\documentclass[prd,aps,a4paper,twocolumn,superscriptaddress,nofootinbib,floatfix]{revtex4-2}
\pdfoutput=1

\usepackage[T1]{fontenc}
\usepackage[utf8]{inputenc}
\usepackage{lmodern}

\usepackage{amsmath,amssymb,amsfonts,bm,mathrsfs}

\usepackage{graphicx}
\usepackage[caption=false]{subfig} 
\usepackage{booktabs}
\usepackage{threeparttable}
\usepackage{dcolumn}
\usepackage[caption=false]{subfig} 

\usepackage{color}
\usepackage{placeins}
\usepackage{float}
\usepackage[normalem]{ulem}
\usepackage{wrapfig}



\usepackage[justification=raggedright]{caption}
\usepackage{braket}
\allowdisplaybreaks
\usepackage[linktoc=none]{hyperref}
\hypersetup{colorlinks,			
	linkcolor=blue,
	citecolor=blue}
\allowdisplaybreaks
\begin{document}

\title{Reconstruction of Black Hole Ringdown Signals with Data Gaps using a Deep-Learning Framework}
\author{Jing-Qi Lai\footnote{E-mail: laijingqi19@mails.ucas.ac.cn}}
\affiliation{School of Physical Sciences, University of Chinese Academy of Sciences, Beijing 100049, China}
\author{Jia-Geng Jiao\footnote{E-mail: jiaojiageng@ucas.ac.cn}}
\affiliation{International Centre for Theoretical Physics Asia-Pacific,
University of Chinese Academy of Sciences, Beijing 100190, China
}
\author{Cai-Ying Shao\footnote{E-mail: shaocaiying@ucas.ac.cn}}
\affiliation{School of Physical Sciences, University of Chinese Academy of Sciences, Beijing 100049, China}

\author{Jun-Xi Shi\footnote{E-mail: shijunxi23@mails.ucas.ac.cn}}
\affiliation{International Centre for Theoretical Physics Asia-Pacific,
University of Chinese Academy of Sciences, Beijing 100190, China
}
\author{Yu Tian\footnote{E-mail:  ytian@ucas.ac.cn}}
\affiliation{School of Physical Sciences, University of Chinese Academy of Sciences, Beijing 100049, China}
\affiliation{Institute of Theoretical Physics, Chinese Academy of Sciences, Beijing 100190, China}

\begin{abstract}
We introduce DenoiseGapFiller (DGF), a deep-learning framework specifically designed to reconstruct gravitational-wave ringdown signals corrupted by data gaps and instrumental noise. DGF employs a dual-branch encoder–decoder architecture, which is fused via mixing layers and Transformer-style blocks. Trained end-to-end on synthetic ringdown waveforms with gaps up to 20\% of the segment length, DGF can achieve a mean waveform mismatch of $0.002$. The residual amplitudes of the Time-domain shrink by roughly an order of magnitude and the power spectral density in the 0.01-1 \,Hz band is suppressed by 1–2 orders of magnitude, restoring the peak of quasi-normal mode(QNM) in the time-frequency representation around 0.01–0.1 \,Hz. 
The ability of the model to faithfully reconstruct the original signals, which implies milder penalties in the detection evidence and tighter credible regions for parameter estimation, lay a foundation for the following scientific work.





\end{abstract}

\maketitle


\section{Introduction}

Gravitational waves (GWs) from compact binary coalescences encode valuable information about the dynamical and strong-field regime of gravity\cite{abbott2016observation}. 
Space-based interferometers such as LISA~\cite{amaroseoane2017laserinterferometerspaceantenna,robson2019construction} are expected to observe higher-SNR signals from supermassive black hole binaries (MBHBs), enabling precision spectroscopy of QNMs in the millihertz band.

The ringdown waveform is well described by a superposition of damped sinusoids known as quasi-normal modes (QNMs), each characterized by a complex frequency $\omega_{lmn} = \omega_R - i \omega_I$ depending solely on the mass $M$ and spin $a$ of the final black hole~\cite{berti2006gravitational}. The fundamental $(\ell, m, n) = (2,2,0)$ mode typically dominates the signal, with a frequency around $f_{\mathrm{QNM}} \sim (1-100)$~mHz for LISA-band black holes with masses in the range $10^5$–$10^7 M_\odot$~\cite{berti2009qnm}. The ringdown amplitude decays exponentially with a timescale $\tau = 1/\omega_I$ typically of order tens to hundreds of seconds for such systems. These clean, short-duration signals serve as natural probes for testing the Kerr nature of astrophysical black holes~\cite{Isi_2019}. 


%



Despite its scientific potential, space-based observation presents unique data integrity challenges. Operational procedures such as antenna repointing, orbital maintenance, and thermal fluctuations can introduce scheduled or unscheduled gaps into the data stream~\cite{baghi2019gaps, dey2021gaps, Spadaro:2023hlw}. These interruptions, even if brief, may erase or corrupt crucial signal segments, particularly the ringdown phase, where both amplitude and phase coherence are essential for black hole spectroscopy. A recent study~\cite{shi2024effectsdatagapsringdown} by Shi \textit{et al.} investigated in detail how such data gaps impact ringdown observability and parameter estimation in space-based joint observation scenarios, highlighting the importance of robust gap recovery methods in this regime. Additionally, gaps result in nonstationarity in the noise spectrum and lead to spectral leakage in Fourier-domain analysis~\cite{edwards2020nonstationary, Nissanke:2010tm, Zaldarriaga:2003ip}. Similar challenges have been studied in the context of ground-based detectors, where non-Gaussian transients (“glitches”) and nonstationary noise artifacts also degrade parameter estimation~\cite{George:2017pmj}.

To address data gaps, several strategies have been explored. Interpolation-based methods
(e.g., cubic splines or linear regression) are simple but can distort narrow-band spectral
content and instantaneous phase, leading to frequency-domain bias in oscillatory signals
\cite{Horowitz1974,Maeland1988,Unser1999}. Windowing mitigates spectral leakage induced
by missing samples, whereas time-delay interferometry (TDI) targets laser-noise
cancellation rather than imputing missing data \cite{Harris1978,Vallisneri2005,Baghi2019}.
Bayesian data augmentation (BDA) treats the unobserved samples as latent variables and
samples them jointly with source parameters, providing statistically consistent gap
handling at the cost of higher computational load and prior sensitivity \cite{Baghi2019}.
Sparse inpainting leverages Fourier/wavelet-domain sparsity and has been adapted to LISA
galactic-binary analyses \cite{Blelly2021,Starck2010}, though coherent, damped sinusoidal
structures (e.g., ringdown QNMs) can remain challenging when gaps coincide with rapid
phase evolution.


Recent studies have proposed deep learning-based gap imputation frameworks to address this issue. Xu \textit{et al.}~\cite{Xu_2024} introduced a DenseNet–BiLSTM architecture that recovers full inspiral–merger–ringdown (IMR) signals in the presence of data corruption. Mao \textit{et al.}~\cite{Mao:2025cae} developed a stacked hybrid autoencoder combining a denoising convolutional autoencoder (DCAE) with a bidirectional gated recurrent unit (BiGRU) decoder, optimized for long-sequence recovery in the context of LISA and demonstrating over 99\% overlap when gaps avoid the merger phase. Wang \textit{et al.}~\cite{Wang:2024waveformer} proposed WaveFormer, a transformer-based denoising pipeline for LIGO data, achieving percent amplitude and phase recovery even in the presence of large glitches.

Though their effectiveness towards their situations, these models are not specifically optimized for the noisy gapped ringdown data, which means a dual objective task including denoising and gap imputation. What's more, the ringdown phase presents unique challenges due to its short duration, high damping, and narrow frequency content. Accurate reconstruction of this segment is crucial for theoretical tests of general relativity based on ringdown analysis, necessitating a gap imputation model specifically for the ringdown phase.

Based on this, we introduce DGF, a deep learning framework specifically designed to reconstruct gravitational wave ringdown signals corrupted by gaps and noise. The model adopts a unified encoder–decoder architecture with Q-transform-based time–frequency inputs~\cite{Chatterji:2004qtf}, multi-channel embeddings that encode statistical context of the missing region, and a TimeMixer-based~\cite{wang2025timemixergeneraltimeseries} core and Transformer-style blocks~\cite{devlin2019bertpretrainingdeepbidirectional} that efficiently model temporal dependencies across masked segments. The entire pipeline is trained end-to-end using a composite loss function. On synthetic ringdown datasets with noise and injected gaps, DGF achieves an average mismatch of $0.002$, peak phase deviations below $0.90^\circ$, and restores spectrogram ridge features characteristic of quasi-normal mode oscillations\cite{7362631}. 

The remainder of this paper is organized as follows: In Sec.~\ref{sec:method}, we give a brief introduction to the background knowledge and then detail the model architecture and training procedure. Sec.~\ref{sec:results} presents quantitative and visual evaluation of DGF under various gap and noise scenarios. We conclude our work and discuss the applicability, limitations, comparsion to existing work and future work in Sec.~\ref{sec:discussion} .


\section{Methodology}
\label{sec:method}

\subsection{Noise Model}
We introduce the LISA sensitivity curve\cite{Robson:2019psd} as the following equations:
\begin{align}
S_n(f) &= \frac{10}{3L^2}
\left[
P_{\rm OMS}(f)
+\frac{4\,P_{\rm acc}(f)}{(2\pi f)^4}\right]
\left[1+\Big(\frac{6}{10}\frac{f}{\,f_\ast}\Big)^2\right] + S_{\rm c}(f),\\
P_{\rm OMS}(f) &=
\left(1.5\times10^{-11}\,{\rm m}\right)^2
\!\left[1+\left(\tfrac{2\,{\rm mHz}}{f}\right)^4\right],\\
P_{\rm acc}(f) &=
\left(3\times10^{-15}\,{\rm m\,s^{-2}}\right)^2
\!\left[1+\left(\tfrac{0.4\,{\rm mHz}}{f}\right)^2\right]
\!\left[1+\left(\tfrac{f}{8\,{\rm mHz}}\right)^4\right].
\label{sensitivity}
\end{align}
where $L=2.5\rm{Gm}$ representing the arm-length of the detector, $f_\ast=c/2\pi L=19.09\rm{mHz}$. The estimated values of the confusion noise using the new LISA design are presented in Ref\cite{Cornish_2017} and are well 
\begin{equation}
S_{\rm c}(f)=
A\,f^{-7/3}\,
\exp\!\big[-f^\alpha+\beta f\sin(\kappa f)\big]\,
\Big[1+\tanh\!\big(\gamma(f_k-f)\big)\Big],
\label{eq:Sconf_refined}
\end{equation}
with coefficients \(\{A,\alpha,\beta,\kappa,\gamma,f_k\}\) chosen for the observation time as tabulated in Ref\cite{Robson:2019psd}.

In addition to the analytic sensitivity adopted above, we benchmark our noise model against an empirical power spectral density constructed from Mock LISA Data Challenge (MLDC/LDC) simulations. Specifically, we estimate PSDs from the LDC noise realizations and interpolate them onto our analysis frequency grid. These empirical PSDs are then used to synthesize colored Gaussian noise consistent with the LISA noise environment, which we inject into the simulated data for subsequent model training and testing.
\subsection{Ringdown Signal}
The ringdown signal can be expressed as a superposition of quasinormal modes (QNMs) of the remnant Kerr black hole. Now we focus on fundamental(n=0) mode, which means given the typical index $(l,m)$, one can obtain the two ringdown polarizations after summing over all$(l,m)$ modes below\cite{Baibhav_2019,Berti_2007}:
\begin{align}
h_{+}^{lm}(t)&=\frac{M_z A_{lm}Y_{+}^{lm}}{r}\mathrm{Re} (e^{-t/\tau_{lm}+i(\omega_{lm}t+\phi_{lm})})\\
h_{\times}^{lm}(t)&=\frac{M_z A_{lm}Y_{\times}^{lm}}{r}\mathrm{Im} (e^{-t/\tau_{lm}+i(\omega_{lm}t+\phi_{lm})}),
\label{eq:ringdown}
\end{align}

where $M_z$ is the redshifted mass of the black hole, $\omega_{lm}$ is the oscillation quasi-normal frequency, $\tau_{lm}$ is the damping time. The angular function can be written as
\begin{align}
Y_{+}^{lm}(\iota)&={}_{-2}Y^{lm}(\iota,0)+(-1)^l {}_{-2}Y^{l -m}(\iota,0)\\
Y_{\times}^{lm}(\iota)&={}_{-2}Y^{lm}(\iota,0)-(-1)^l {}_{-2}Y^{l -m}(\iota,0),
\label{eq:angular function}
\end{align}
and $\omega_{lm}$, $\tau_{lm}$ in Eq.~\eqref{eq:ringdown} can be given by the following expression\cite{Berti_2016}:
\begin{align}
\omega_{lm}&=\frac{f_1+f_2(1-\chi_f)^{f_3}}{M_z}\\
\tau_{lm}&=\frac{2(q_1+q_2(1-\chi_f)^{q_3})}{\omega_{lm}}.
\label{eq:omega and tau}
\end{align}
The fitting parameters in the expression are shown in Table \ref{fitting para}. For remnat black hole without spin, $\chi_f$ can be calculated as\cite{Barausse_2009}
\begin{align}
    \chi_f(q)=\eta(2\sqrt{3}-3.517\eta+2.5763\eta^2),
\label{eq:angular momentum}
\end{align}
where $q$ is the mass ratio and $\eta=q/(1+q)^2$ is the symmetric mass ratio.

\begin{table}[t]

  \caption{The fitting parameters for Eq.~\eqref{eq:omega and tau}. Taken from Ref\cite{Berti_2006}.}
  \label{tab:fit-coeff}
  \begin{ruledtabular}
    \begin{tabular}{crrrrrr}
      $(\ell,m)$ & $f_1$   & $f_2$    & $f_3$   & $q_1$   & $q_2$   & $q_3$   \\
      \hline
      $(2,2)$    & 1.5251  & -1.1568  & 0.1292  & 0.7000  & 1.4187  & -0.4990 \\
      $(3,3)$    & 1.8956  & -1.3043  & 0.1818  & 0.9000  & 2.3430  & -0.4810 \\
      $(2,1)$    & 0.6000  & -0.2339  & 0.4175  & -0.3000 & 2.3561  & -0.2277 \\
      $(4,4)$    & 2.3000  & -1.5056  & 0.2244  & 1.1929  & 3.1191  & -0.4825 \\
    \end{tabular}
  \end{ruledtabular}
 \label{fitting para}
\end{table}
\subsection{Data Gaps}
\label{subsec:data_gaps_condensed}
Space-based detectors inevitably feature scheduled gaps (e.g., antenna repointing, test-mass discharging) and unscheduled gaps (anomalies/glitches). In mission-level requirements, duty cycles $\gtrsim 75\%$ are typically assumed for worst-case assessments \cite{AmaroSeoane:2021}. On long baselines, even short missing stretches can compromise stationarity and phase coherence, and---being equivalent to time-domain windowing---cause narrow-band signal power to leak into side lobes in the Fourier domain \cite{Harris:1978}.

A simple yet effective stochastic model treats the waiting time $\Delta T$ between neighboring gaps as exponentially distributed,
\begin{equation}
p(\Delta T)=\lambda\,e^{-\lambda \Delta T},
\end{equation}
with the rate $\lambda$ chosen to satisfy the target duty cycle. To synthesize gaps in strain data $h(t)$, a window $G(t)$ is applied,
\begin{equation}
h_G(t)=G(t)\,h(t),
\end{equation}
where $G(t)=0$ inside the gap and $G(t)=1$ otherwise.

Previous studies find that scheduled gaps have modest effects, while random (unscheduled) gaps can markedly degrade detectability and widen posteriors, especially when a gap overlaps fast-evolving phases.

\subsection{Model Analysis}
\begin{figure}[htbp]
    \includegraphics[width=0.48\textwidth]{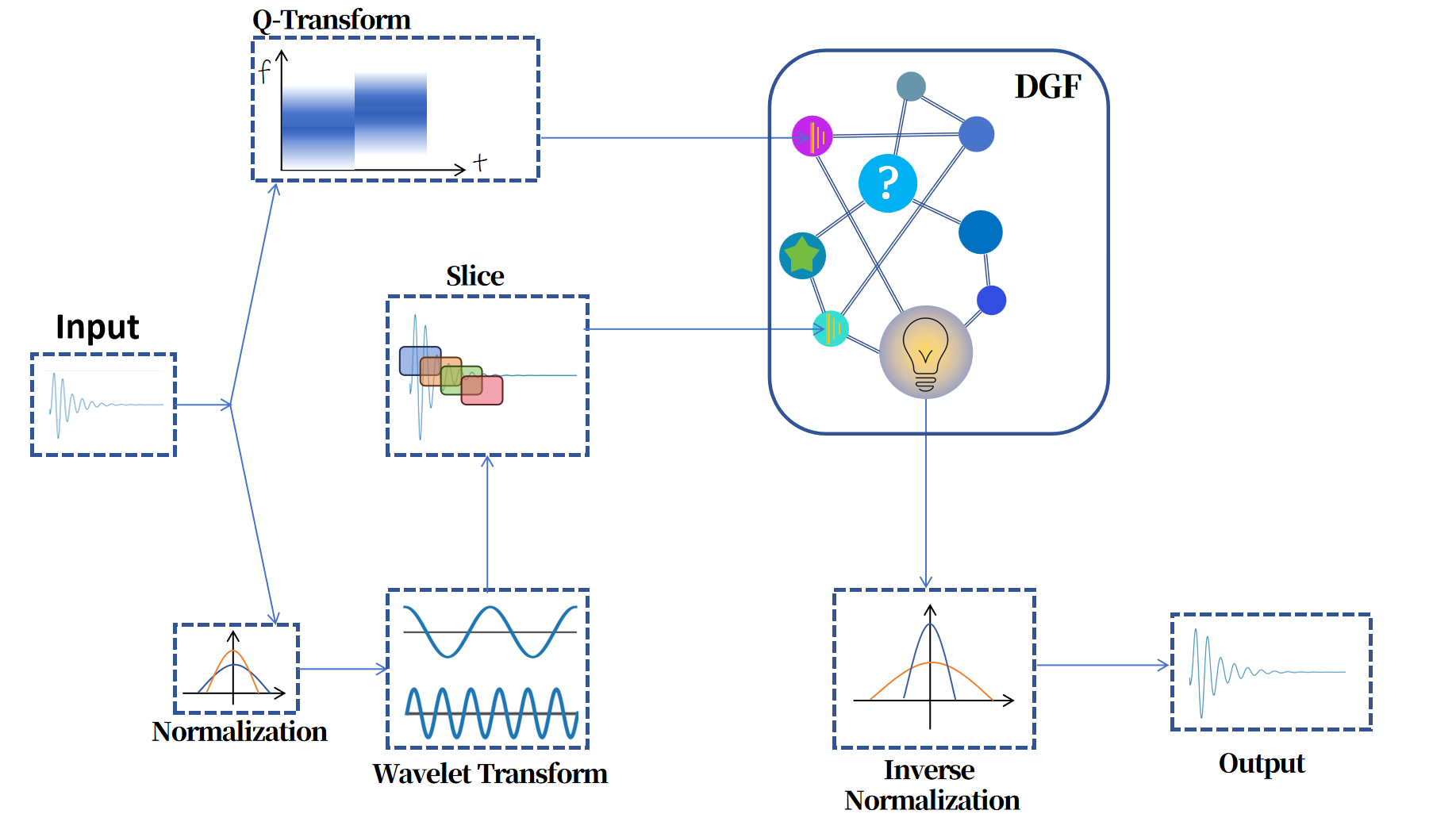}
    \caption{\textbf{Schematic overview of the pipeline of DGF.} The input data with a length of 1056 (noise plus signal with gap) goes through two branches. One branch directly performs q-transform on the data to obtain the amplitude and angle in the time direction and frequency scale, which is used as the first receiver of the dual-channel image format data input into the model, and the other branch normalizes data and performs wavelet transform processing to obtain the data of 8 channels. Finally, the data of each channel is segmented with 50\% overlap to obtain a 32-group signal with a length of 64. Each group is regarded as a token input into the second receiver of the model. Output from DGF model is inverse-normalized to obtain the reconstructed waveform.}
    \label{fig:pipeline}
\end{figure}

\begin{figure}[htbp]
    \centering
    \includegraphics[width=0.48\textwidth]{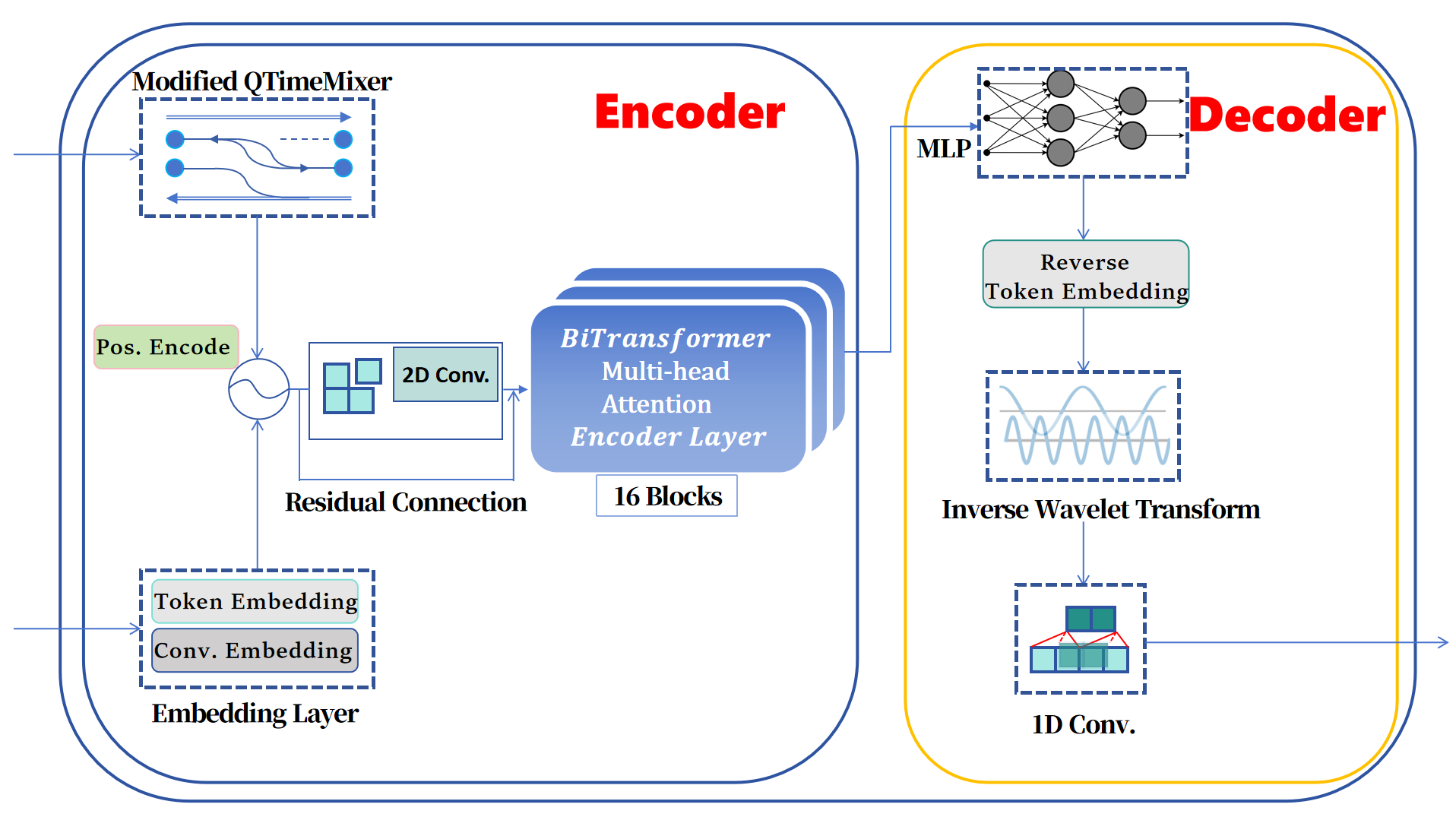}
    \caption{\textbf{Schematic overview of the DGF architecture.} The dual-branch encoder receives Q-transform and tokenized wavelet patches in parallel. The embedding module integrates token, convolutional embeddings to encode localized time–frequency features and statistical context. Positional embeddings are added, followed by residual connected 2D convolutions and stacked BiTransformer encoder layers (16 blocks) with multi-head attention to model both short- and long-range dependencies. The decoder contains an MLP layer, followed by reverse token embedding and inverse wavelet transform and 1D convolution to reconstruct the final waveform.}
    \label{fig:architecture}
\end{figure}

We propose DGF, an encoder–decoder-based deep learning architecture designed to reconstruct ringdown waveforms from gravitational wave time-series data containing nontrivial gaps and noise. The core design is motivated by the need to preserve physical fidelity in both amplitude and phase, and to respect the localized and damped nature of QNM signals.
As shown in Fig.~\ref{fig:pipeline} and Fig.~\ref{fig:architecture}, DGF overall pipeline consists of three main components: (1) a preprocessing flow of the dual-branch (2) a hybrid Encoder module combining convolutional, and Token learning and a sequence modeling network built from stacked TimeMixer blocks before positional encoding and Transformer block layers with residual connection (3) a relatively simple Decoder with MLP layers and reverse embedding and inverse transform and a 1D convolution. The final output reconstruction appears to be a complete waveform after inverting the normalization applied during preprocessing.

Compared to recent deep learning methods for data gap imputation, our framework is specifically tailored for the ringdown phase. The Q-transform and wavelet transformation enhance the visibility of ridge structures, facilitating localized context encoding around the gap\cite{Chatterji:2004qtf, McNabb:2019spectrogram} and extracting the characteristics of the noise. While both the constant Q-transform and discrete wavelet transform (DWT) are time–frequency representations capable of capturing nonstationary signal components, they differ in formulation, resolution, and application focus. The Q-transform~\cite{Chatterji:2004qtf} is based on short-time Fourier transform with a logarithmic frequency spacing and constant quality factor $Q = f/\Delta f$, allowing it to capture long-lived, narrowband structures such as the 
QNM ridges in black hole ringdowns. In contrast, the wavelet transform~\cite{MALLAT199942} decomposes signals via scaled and shifted versions of a mother wavelet, which is born to recognize the transient information of the signal, providing convenience for denoising.


Supplemented by two-dimensional convolution processing of the image after the Q-transform, the frequency domain information of the signal is further extracted\cite{younesi2024comprehensivesurveyconvolutionsdeep}. Time-mixing supports efficient and scalable modeling across short-duration sequences~\cite{wang2025timemixergeneraltimeseries}. Furthermore, DGF leverages Transformer-based attention mechanisms to capture long-range dependencies and subtle temporal patterns~\cite{vaswani2023attentionneed,devlin2019bertpretrainingdeepbidirectional}. 


\subsection{Sequence Modeling}

Given a noisy gravitational wave strain sequence containing a masked gap region as input $d=n(t)+G(t)(h(t))$ , where $n(t)$ denotes noise and $G(t)(h(t))$ denotes signal with gaps, the DGF model encodes and decodes the input before output a clean signal waveform $h(t)$. The  framework consists of a dual-branch encoder, a transformer-based sequence modeling core, and a structured decoder which handle the data. It is designed to reconstruct gravitational wave ringdown signals corrupted by gaps and noise using a hybrid representation of time--frequency and wavelet-transformed inputs.\cite{Wang:2024waveformer}

\paragraph{Dual-branch Input Encoding.} Let $x \in \mathbb{R}^{T}$ be the raw 1D time-domain input signal of length $T$. We define two parallel encoding branches:

\begin{enumerate}
    \item \textbf{Q-transform branch:} A Q-transform $\mathcal{Q}: \mathbb{R}^{T} \to \mathbb{R}^{H \times W \times C}$ is applied to obtain a time–frequency representation $X^{(1)} = \mathcal{Q}(x)$, where $H$ and $W$ denote the number of frequency bins and time steps and $C$ the number of channels (amplitude and phase). The output is passed through a modified TimeMixer module $\mathcal{T}$~\cite{wang2025timemixergeneraltimeseries}:
    \begin{equation}
        Z^{(1)} = \mathcal{T}(X^{(1)}),
    \end{equation}
    where $Z^{(1)}$ encodes short-duration temporal and spectral ridge structures in the ringdown signal.  
    \item \textbf{Wavelet branch:} A normalized version of $x$ is transformed via a discrete wavelet transform (DWT) $\mathcal{W}$ into multi-resolution coefficients $x_{w} = \mathcal{W}(x) \in \mathbb{R}^{C \times T}$, where $C$ is the number of wavelet channels~\cite{MALLAT199942}. The coefficients are segmented into $N$ overlapping chunks (tokens) of length $L$ with stride $L/2$ to form a token sequence\cite{nie2023timeseriesworth64}:
    \begin{equation}
        X^{(2)} = [x_{w}^{(1)}, x_{w}^{(2)}, \dots, x_{w}^{(N)}], \quad x_{w}^{(i)} \in \mathbb{R}^{C \times L}.
    \end{equation}
    Each token is embedded as $z_{i}^{(2)} = \mathcal{E}(x_{w}^{(i)}) \in \mathbb{R}^{d}$, where $\mathcal{E}$ includes token embedding and convolutional embedding blocks~\cite{Mao:2025cae, Xu_2024}. The sequence $Z^{(2)} = [z_1^{(2)}, \dots, z_N^{(2)}] \in \mathbb{R}^{N \times d}$ is then fused with positional encoding~\cite{vaswani2023attentionneed}.  
\end{enumerate}

\paragraph{Transformer Encoder.} The concatenated representation $Z = Z^{(1)} \oplus Z^{(2)}$ enters into a 2D convolution with residual connection and then be passed through stack of 16 bidirectional multi-head attention transformer blocks:
\cite{devlin2019bertpretrainingdeepbidirectional,xiong2020layernormalizationtransformerarchitecture}

\begin{align}
    Z_0' &= Z + \text{PosEmbed}(Z) \\
    Z_0  &= \text{Conv2d}(Z_0')+\text{Res}(Z_0')\\
    Z_{\ell} &= \text{TransformerBlock}_{\ell}(Z_{\ell-1}), \quad \ell=1,\dots,16,
\end{align}
where PosEmbed denotes positional embedding, Conv2d denotes 2D convolution, and Res denotes residual connection, and each TransformerBlock consists of:
\begin{itemize}
    \item Multi-head self-attention: captures long-range dependencies
    \item Feedforward layer: non-linear representation learning
    \item Residual connections and normalization for stability
\end{itemize}

\paragraph{Decoder.} The encoded output $Z_{16}$ is processed through:
\begin{align}
  Y &= \operatorname{MLP}(Z), \\
  \hat{x}_{\mathcal E'} &= (\mathcal E')^{-1}(Y), \\
  \hat{x}_{w} &= \mathcal W^{-1}\!\bigl(\hat{x}_{\mathcal E'}\bigr), \\
  \hat{x} &= \operatorname{Conv1d}\!\bigl(\hat{x}_{w}\bigr).
\end{align}

where MLP denotes multilayer perceptron, $\mathcal{E}^{\prime-1}$ denotes the reverse token embedding and $\mathcal{W}^{-1}$ is the inverse wavelet transform and $\text{Conv1d}$ denotes the 1D convolution. Finally, inverse normalization is applied to obtain the reconstructed time-domain signal $\hat{x}$.



\subsection{Loss Function and Training Strategy}

The DGF model is trained to directly reconstruct the complete gravitational waveform from input sequences containing additive noise and artificially masked gap regions. The loss function used is the standard mean squared error (MSE) that computed over the entire sequence plus a L1 regularization term:
\begin{equation}
\mathcal{L}_{\text{Total}}
= \mathcal{L}_{\text{MSE}} + \mathcal{L}_{\text{Reg}}
= \frac{1}{T}\sum_{t=1}^{T}\!\bigl(\hat{h}(t)-h(t)\bigr)^{2}
  + \alpha \sum_{i}\!\left\lvert \theta_{i} \right\rvert .
\end{equation}

where $\alpha$ is the weight of L1 regularization term which is increasing as training process, and $\theta$ is the parameters of model, and $h(t)$ is the clean ground-truth waveform and $\hat{h}(t)$ is the model output. Unlike some reconstruction frameworks that restrict loss evaluation to observable (non-gap) regions~\cite{mao2024bigru}, our approach imposes a global supervision objective, encouraging the network to learn both denoising and gap inpainting simultaneously.\cite{zhang2018unreasonableeffectivenessdeepfeatures}


The training dataset consists of synthetic ringdown signals from parameters space of 
(Total Mass, Mass Ratio, Redshift)=$\left(\mathrm{1e5-1e6M_{\odot}, 0.5-0.8, 4-10}\right)$, 
corrupted by additive Gaussian noise with LISA power spectral density (PSD) and zeroed-out segments (gaps) of variable duration up to 20\% ringdown signal length, typically 100 points within a 1056-sample sequence. The model is trained to recover the clean signal across the entire sequence, including the missing portions. As mentioned earlier, the ringdown signal waveform model is based on Ref\cite{zhang2021parameter} and the noise data are interpolation simulation of LISA Data Challenge(LDC) \cite{Babak_2008,Babak_2010}.

Optimization is performed using the Adam algorithm with an initial learning rate of $3\times10^{-4}$ and batch size of 32. Early stopping is based on validation loss. Despite the simplicity of the loss function, we observe that the model is able to restore spectral and phase structure with high fidelity, as evaluated in Section~\ref{sec:results}.

\subsection{Mean Squared Error and Mean Absolute Error}\label{subsec:MeanSquaredError}
To quantify the amplitude reconstruction accuracy, we compute the Mean Squared Error (MSE) and Mean Absolute Error (MAE) between the DGF final output \(\hat{h}(t)\) and the ground truth \(h(t)\) on SNR-partitioned test subsets. The MSE and MAE is defined as
\begin{equation}
\mathrm{MSE} = \frac{1}{N} \sum_{i=1}^{N} \bigl[h_i - \hat{h}_i\bigr]^2,
\end{equation}
\begin{equation}
    \mathrm{MAE}=\frac{1}{N} \sum_{i=1}^{N}\left\lvert h_i - \hat{h}_i \right\rvert
\end{equation}
where \(N\) is the number of time samples.

\begin{figure}[htbp]
  \centering
  \subfloat[\textbf{lower-SNR 1--5}: mean MSE $=3.01\times10^{-2}$ and mean MAE $=1.91\times10^{-1}$.%
  \label{fig:mse_1-5}]{
    \includegraphics[width=0.48\textwidth]{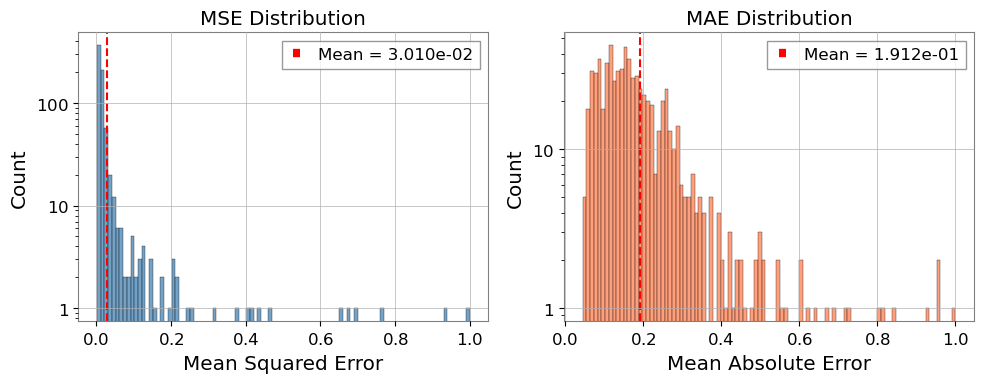}
  }\hfill
  \subfloat[\textbf{higher-SNR 5--10}: mean MSE $=2.12\times10^{-2}$ and mean MAE $=1.51\times10^{-1}$.%
  \label{fig:mse_5-10}]{
    \includegraphics[width=0.48\textwidth]{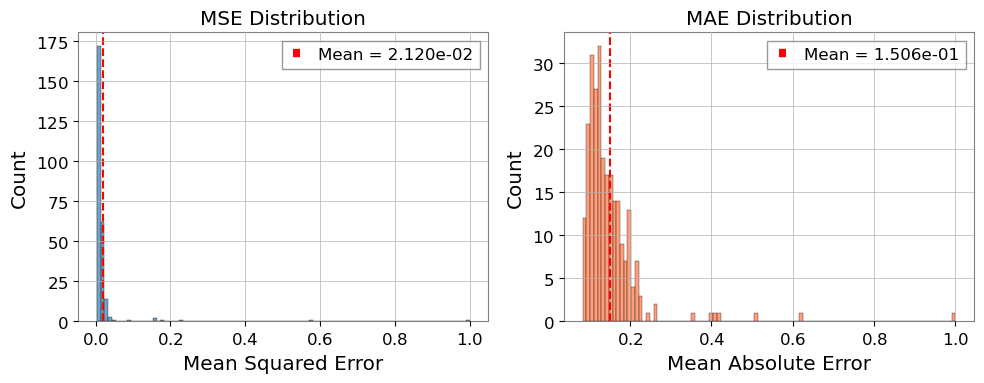}
  }
  \caption{\textbf{Histograms of MSE and MAE for the two SNR regimes.} Improved amplitude recovery at higher SNR.}
  \label{fig:mse_snr}
\end{figure}

To avoid numerical accuracy errors and to observe the distribution more intuitively, we scale the output and the actual amplitude, that is, multiply by 1e20, which is the typical value of GWs. Fig.~\ref{fig:mse_snr} presents the histograms of MSE and MAE for the lower-SNR (1–5) and higher-SNR (5–10) groups. 

In the lower-SNR subset (Fig.~\ref{fig:mse_1-5}), the MSE distribution is tightly clustered with a mean of \(3.010\times10^{-2}\) and for MAE that is \(1.912\times10^{-1}\). In the 5–10 subset (Fig.~\ref{fig:mse_5-10}), the mean MSE makes slight improvement to \(2.12\times10^{-2}\) and mean MAE is \(1.506\times10^{-1}\). 

Building on our statistical analysis of the per-sample MSE distribution, we now turn to the training and validation loss curves to more clearly illustrate how the model converges and generalizes over successive epochs.

The loss curves exhibit a steep decline in both training and validation loss during the early epochs as the network quickly learns to denoise and reconstruct basic waveform signal.  After epoch 10, the rate of decrease slows, and after that the loss value continuously declines steadily and by epoch 90 both curves reach a stable minimum of 50.  The close alignment of validation loss with training loss throughout—and the absence of a widening gap—suggests that the model is not overfitting and maintains robust performance on unseen data.  This convergence behavior confirms that the chosen architecture and optimization schedule effectively balance learning capacity and regularization. 

\begin{figure}[htbp]
  \centering
  \includegraphics[width=0.48\textwidth]{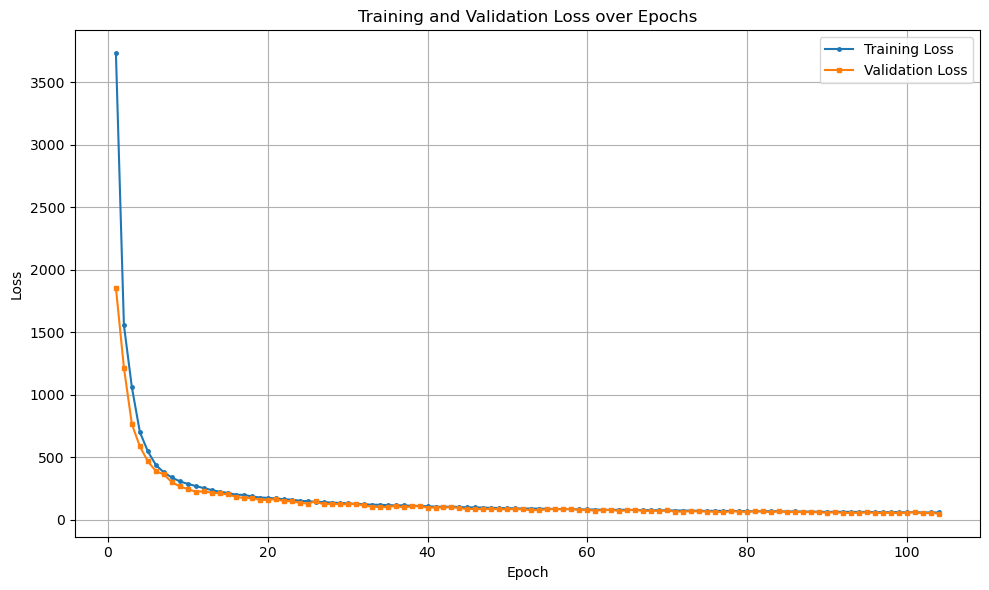}
  \caption{\textbf{Training (blue) and validation (orange) loss curves for the gap-filling model.} Both losses decrease rapidly during the first 10 epochs and then plateau, with validation closely tracking training, indicating good generalization and convergence.}
  \label{fig:loss}
\end{figure}

Sec.~\ref{subsec:Mismatch} and \ref{subsec:MeanSquaredError} represent clear SNR dependence demonstrates that higher-SNR events are reconstructed with markedly greater fidelity, achieving highly matching in the ringdown segment. We require the DGF model to have high-precision reconstruction capabilities for higher-SNR events, and at the same time, we also need the DGF to have good feedback for lower-SNR. Despite the fact that the mismatch of the DGF reconstruction signal for events with higher-SNR is significantly lower, which is reasonable, we also found that the reconstruction for events signal with lower-SNR has very satisfactory results.

\section{Results}
\label{sec:results}
We evaluate the performance of the DGF model on synthetic gravitational wave datasets simulating the ringdown phase of binary black hole mergers. Each sample consists of a time-domain waveform generated as a sum of damped sinusoids, representing QNM oscillations, with additive colored Gaussian noise consistent with the LISA PSD\cite{Berti:2022kcw,Robson_2019}. The signals are randomly sampled in mass, spin, and orientation parameters to introduce variability in frequency and damping time\cite{berti2006gravitational,Khan_2016,Husa_2016}.

To demonstrate the generalization performance of the model and its behavior for different SNR\,\cite{robson2019construction}, we divide the test set into two parts based on SNR\,. One part has SNR\, ranging from 1 to 5, and the other part has SNR\, ranging from 5 to 10.

For the SNR\,5–10 test set, the DGF model reconstructs the noisy signal with gap well: across different QNM oscillation periods and varying gap positions, the reconstructed waveform (green) closely follows the true waveform (blue), with only minimal deviation at the noise floor. At the edge of the gap, DGF output smoothly bridges the missing region, preserving the oscillation pattern and damping rate. Notably, the reconstructed waveform aligns closely with the ground truth not only in amplitude but also in phase, without introducing noticeable artifacts at the gap boundary. This reflects the model’s ability to infer the underlying QNM mode continuation using global and local context. Even for gaps located near the merger peak, the model retains high fidelity. 

\begin{figure}[htbp]
  \centering
  \subfloat[]{
    \includegraphics[width=0.48\textwidth]{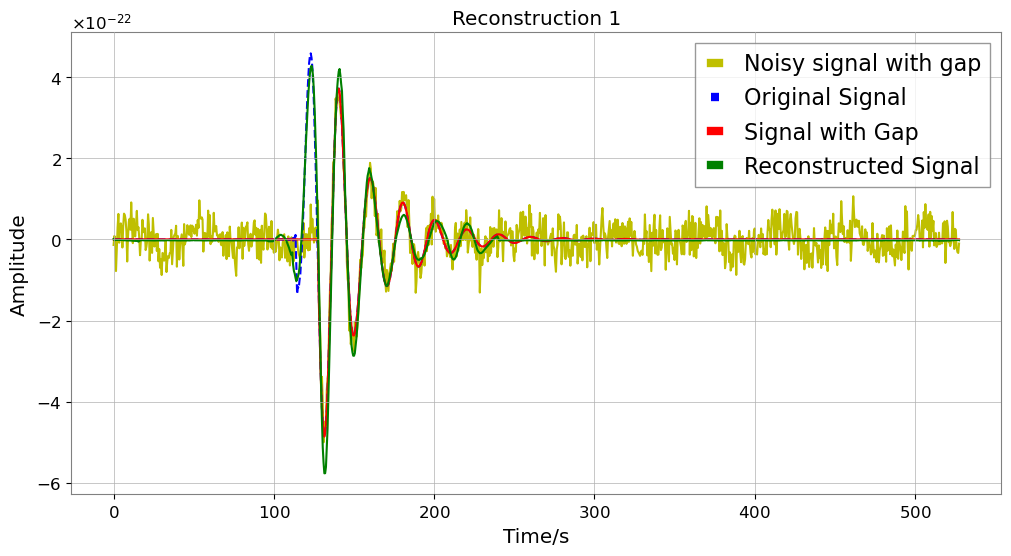}
    \label{fig:glance1}
  }\hfill
  \subfloat[]{
    \includegraphics[width=0.48\textwidth]{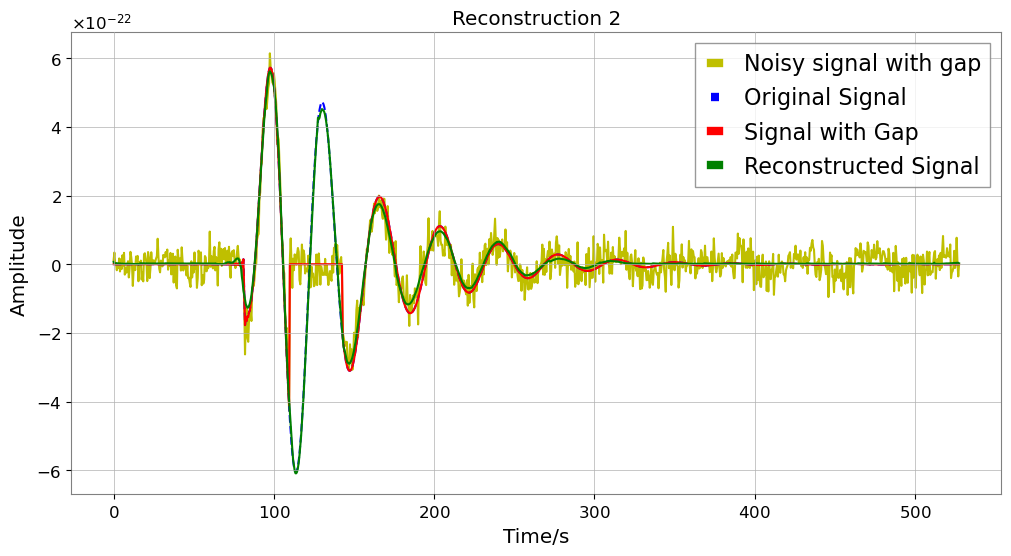}
    \label{fig:glance2}
  }\hfill
  \subfloat[]{
    \includegraphics[width=0.48\textwidth]{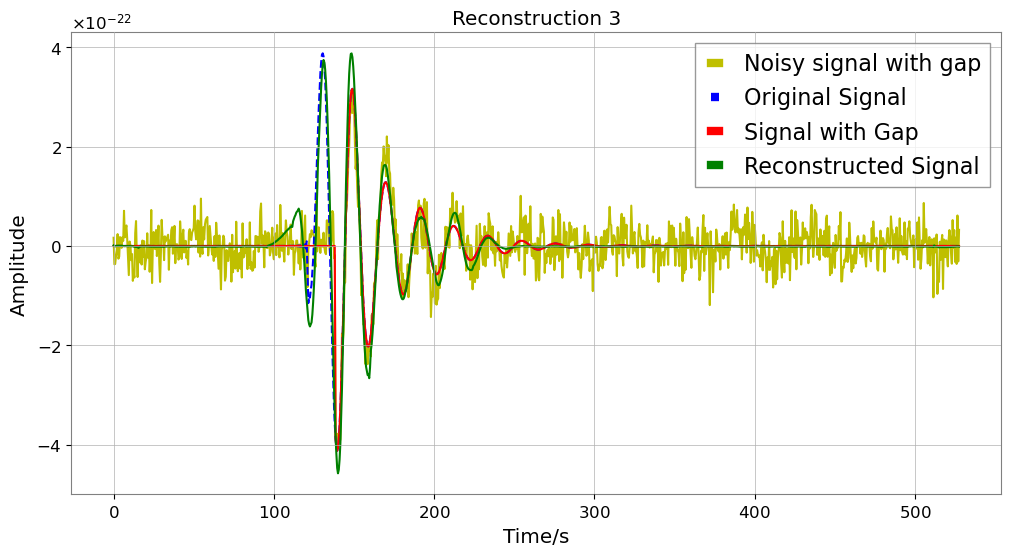}
    \label{fig:glance3}
  }
  \caption{\textbf{Qualitative examples of DGF reconstruction on test samples.} The green trace is the reconstructed signal, red indicates the gap region, yellow is the noisy input with gap, and blue is the original signal. The sampling frequency is $2\,\mathrm{Hz}$.}
  \label{fig:glance}
\end{figure}

We also observe that in lower-SNR cases where the gap covers a majority of the ringdown content, DGF can still recover the qualitative structure of the signal, exhibiting robust noise resilience, nearly fully reconstructing both amplitude and phase around the main peak and partially restoring subsequent oscillations. While small deviations are visible, the waveform maintains the expected damping envelope and frequency content. Fig.\ref{fig:glance} are three representative outputs as example.

These qualitative examples illustrate the DGF model’s ability to recover ringdown waveforms under realistic noise and gap conditions. We now turn to the quantitative evaluation in the following sections.

\subsection{Mismatch}\label{subsec:Mismatch}
To assess reconstruction fidelity, we compute the normalized waveform mismatch \cite{Cutler_1994} and further divide the test set by network signal-to-noise ratio (SNR) into two groups: lower-SNR (1–5) and higher-SNR (5–10) samples\cite{Robson:2019psd}. The  definition of mismatch can be expressed as \ref{formula:mismatch}:

\begin{equation}
\label{formula:mismatch}
\mathcal{M} = 1 - \frac{\langle h_1, h_2 \rangle}{\sqrt{\langle h_1, h_1 \rangle \, \langle h_2, h_2 \rangle}},
\end{equation}
where
\begin{equation}
\langle h_1, h_2 \rangle = 4 \, \mathrm{Re} \int_{f_{\mathrm{low}}}^{f_{\mathrm{high}}} \frac{\tilde{h}_1(f) \, \tilde{h}_2^*(f)}{S_n(f)} \, df.
\end{equation}

\begin{figure}[htbp]
  \centering
  \subfloat[\textbf{Histogram of mismatch values for test samples with network SNR in the range $[1,\,5]$.}
            The DGF model achieves a mean mismatch of $\approx 0.023$ and a median of $0.0049$, indicating robust performance under lower-SNR conditions.]{
    \includegraphics[width=0.48\textwidth]{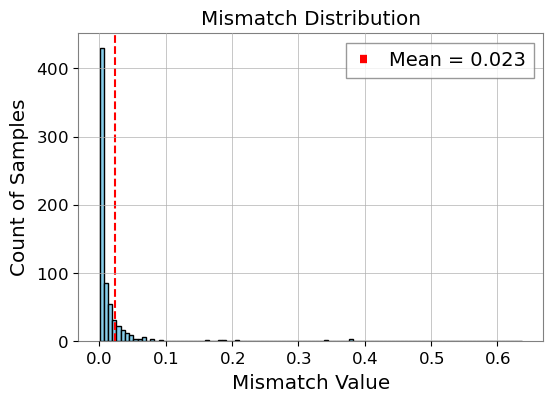}
    \label{fig:mismatch_SNR1-5}
  }\hfill
  \subfloat[\textbf{Histogram of mismatch values for test samples with network SNR in the range $[5,\,10]$.}
            The DGF model reduces the mean mismatch to $\approx 0.002$ and the median to $0.000966$, reflecting substantially improved reconstruction fidelity at higher SNR.]{
    \includegraphics[width=0.48\textwidth]{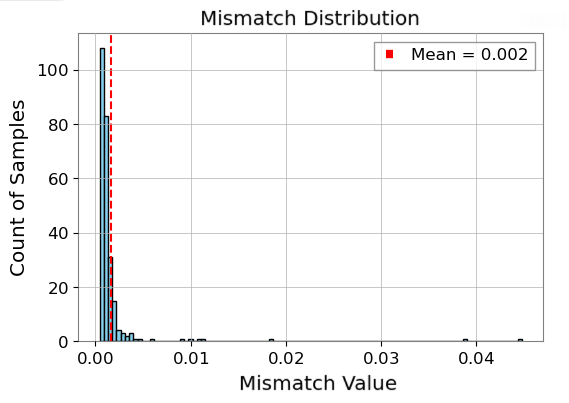}
    \label{fig:mismatch_SNR5-10}
  }
  \caption{}
\end{figure}

Fig.~\ref{fig:mismatch_SNR1-5}, \ref{fig:mismatch_SNR5-10} and Table~\ref{table:mismatch} show the mismatch distributions for each SNR bin. For lower-SNR, the DGF model yields a mean mismatch of approximately 0.023, with a median of 0.0049, 75th percentile of 0.015. Among the 719 samples, more than 95\% of the mismatch between the reconstructed signal and the original signal was below 0.091. There were several cases with a large mismatch, which was speculated to be caused by a too small SNR or extremely small instability of the model. In contrast, for SNR 5–10 the mean mismatch drops to about 0.00246, with a median of 0.000966, 75th percentile of 0.001394, and values ranging from a minimum of 0.000445 and a maximum of 0.044836. 
\begin{table}[ht]
\centering
\caption{Mismatch Statistics}
\label{table:mismatch}
\begin{tabular}{lrr}
\toprule
  & SNR 1–5   & SNR 5–10  \\
\midrule
count   & 719.000   & 258.000   \\
mean    & 0.022808  & 0.002457  \\
std     & 0.059749  & 0.006555  \\
min     & 0.000531  & 0.000375  \\
25\%    & 0.002073  & 0.000754  \\
50\%    & 0.005013  & 0.000966  \\
75\%    & 0.016818  & 0.001439  \\
\bottomrule
\end{tabular}
\end{table}

Beyond serving as a summary statistic of reconstruction fidelity, the mismatch has a direct
connection to detection strength, model selection, and parameter inference under the conventional
noise-weighted inner product. Let $\mathcal{O}\equiv 1-\mathcal{M}$ denote the (noise–weighted) overlap
between the recovered signal and the true waveform using the same noise PSD $S_n(f)$ as in
Eq.~\eqref{formula:mismatch}. In matched-filtering with a fixed $S_n(f)$, the achievable SNR with a
template $h_2$ against data containing $h_1$ scales approximately as
\begin{equation}
\rho_{\rm rec}\simeq \mathcal{O}\,\rho_{\rm opt},
\label{eq:snr-vs-mismatch}
\end{equation}
where $\rho_{\rm opt}=\sqrt{\langle h_1,h_1\rangle}$ is the optimal SNR for a perfectly matched template
(e.g., \cite{Owen:1996PRD,Cutler_1994}). Thus a mismatch $\mathcal{M}$ corresponds to a fractional SNR
loss of $\simeq \mathcal{M}$ to first order. Because detection odds depend exponentially on SNR,
a convenient back-of-the-envelope relation for the Bayes factor between a signal+noise model and
noise-only is $\ln\mathcal{B}\propto \rho^2/2$ in the Gaussian-noise, higher-SNR limit
(e.g., \cite{Thrane:2019PASA,Veitch:2015PRD}). Combining this with Eq.~\eqref{eq:snr-vs-mismatch} yields
\begin{equation}
\Delta\ln\mathcal{B}\;\equiv\;\frac{1}{2}\big(\rho_{\rm opt}^2-\rho_{\rm rec}^2\big)
\;\approx\;\frac{1}{2}\,\rho_{\rm opt}^2\big[1-(1-\mathcal{M})^2\big]
\;\simeq\;\mathcal{M}\,\frac{\rho_{\rm opt}^2}{2},
\end{equation}
so even a few-percent mismatch can produce a noticeable decrease in model evidence at moderate
SNR.

For parameter estimation, the Fisher information matrix
$\Gamma_{ij}=(\partial_{\theta_i}h\,|\,\partial_{\theta_j}h)$ implies that posterior credible widths
scale as $\mathrm{Cov}\sim\Gamma^{-1}\propto\rho^{-2}$ when the likelihood is close to Gaussian in the
parameters \cite{Cutler_1994}. Therefore, via Eq.~\eqref{eq:snr-vs-mismatch}, a smaller
mismatch (larger overlap) tightens the credible regions for quantities of interest in ringdown
spectroscopy—such as $(\omega_R,\tau)$ and mode amplitudes $A_{\ell mn}$—while also reducing potential
systematic biases that arise when the residual $r=h_1-h_2$ has non-negligible projections along
$\partial_{\theta_i}h$. In short, the observed distributions in
Figs.~\ref{fig:mismatch_SNR1-5}–\ref{fig:mismatch_SNR5-10} and Table~\ref{table:mismatch} translate,
under fixed $S_n(f)$, into (i) near-optimal SNR recovery in the higher-SNR bin and percent-level SNR
loss in the lower-SNR bin; (ii) correspondingly milder penalties in $\ln\mathcal{B}$; and (iii) tighter,
less biased constraints on $(\omega_R,\tau,A_{\ell mn})$, consistent with the physical improvements
seen in phase continuity and time–frequency ridge sharpness.

\subsection{Amplitude Distribution and Power Spectral Density}\label{subsec:AmplitudeDensityFrequencySpectrum}
\begin{figure}[htbp]
  \centering
  \subfloat[Residual amplitude distribution (SNR 1--5).\label{fig:amp_stat_1-5}]{
    \includegraphics[width=0.45\textwidth]{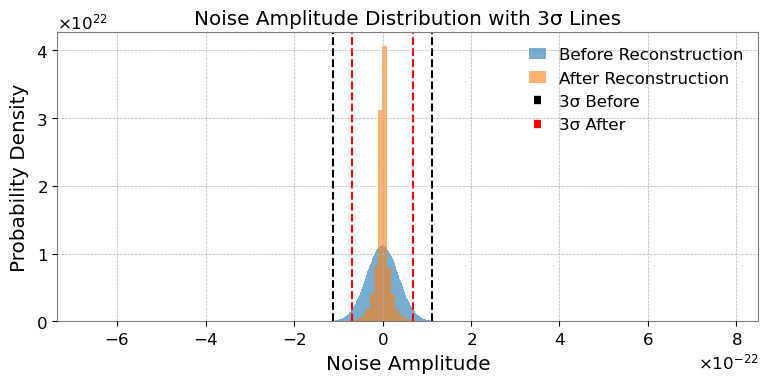}
  }\hfill
  \subfloat[PSD comparison (SNR 1--5).\label{fig:psd_1-5}]{
    \includegraphics[width=0.45\textwidth]{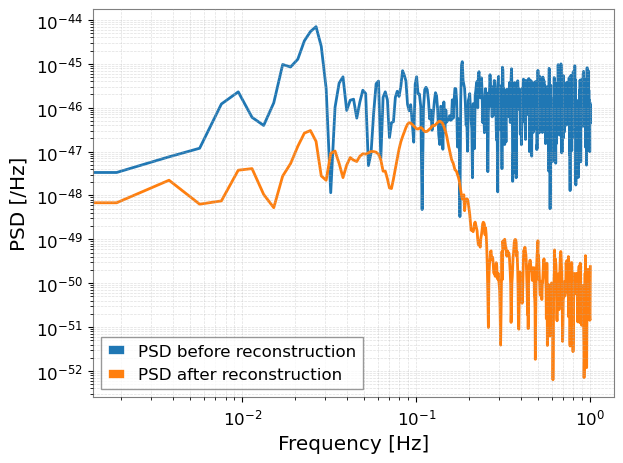}
  }

  \vspace{1em}

  \subfloat[Residual amplitude distribution (SNR 5--10).\label{fig:amp_stat_5-10}]{
    \includegraphics[width=0.45\textwidth]{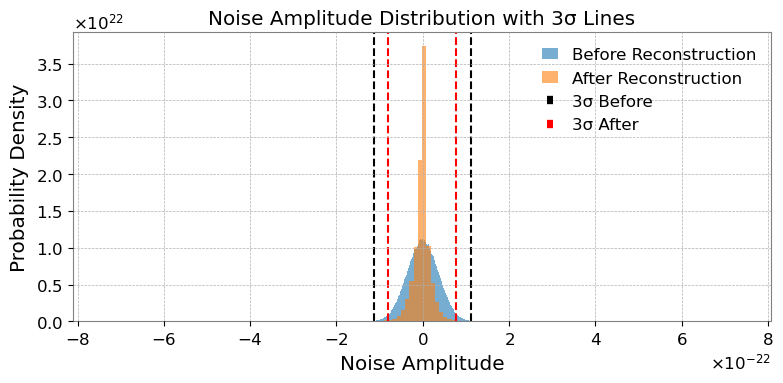}
  }\hfill
  \subfloat[PSD comparison (SNR 5--10).\label{fig:psd_5-10}]{
    \includegraphics[width=0.45\textwidth]{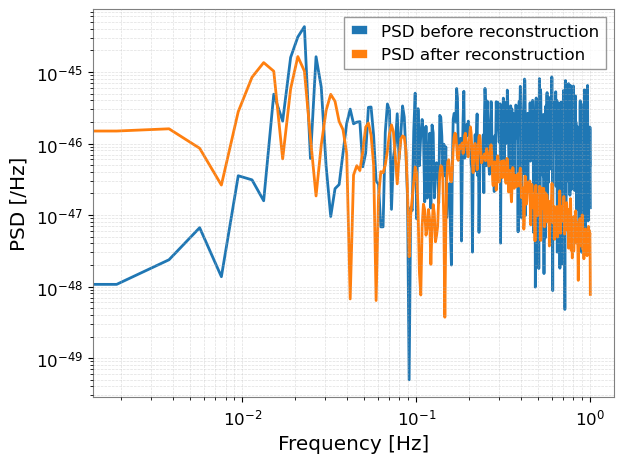}
  }

  \caption{\textbf{Time-domain residual amplitude histograms and frequency-domain PSD before (blue) and after (orange) DGF reconstruction for two SNR regimes.} The model sharpens the residual amplitude distribution around zero and restores the QNM spectral peak while suppressing broadband noise.}
  \label{fig:amp_spec_snr}
\end{figure}

To further evaluate the denoising capability of the DGF model across different SNR regimes, we analyze the residual amplitude distribution and PSD for both lower-SNR (1–5) and higher-SNR (5–10) subsets, where the residual is defined as $h_{\rm res}\equiv h_{\rm recon}-h_{\rm true}$. Figures~\ref{fig:amp_stat_1-5} and \ref{fig:amp_stat_5-10} show the time-domain noise-amplitude histograms before (blue) and after (orange) reconstruction for the lower-SNR subset: the input noise is approximately Gaussian with standard deviation $\sigma \approx 1.12\times 10^{-22}$, while the DGF output concentrates sharply around zero with $3\sigma \approx 7\times 10^{-23}$. Figures~\ref{fig:psd_1-5} and \ref{fig:psd_5-10} present a comparison of the PSDs of the DGF residual noise and the LISA noise model for SNR $1\text{--}5$ and $5\text{--}10$. From these panels one sees that, in the $0.01\text{--}1\,\mathrm{Hz}$ band, the noise power spectral density is suppressed by roughly one to two orders of magnitude; this band is precisely where the dataset’s ringdown power is concentrated (see below). These results confirm the model’s ability to jointly suppress unstructured noise and recover physically meaningful QNM features across varying SNR conditions~\cite{abbott2016observation,Baghi_2019}.

\subsection{Spectral Fidelity and Time–Frequency Coherence}
\label{subsec:SpectralFidelityTimeFrequencyCoherence}

Beyond time–domain accuracy, it is crucial that gap-filling methods recover the ringdown’s narrow-band spectral structure \cite{zhang2018unreasonable,abbott2016observation}. We therefore compute Q-transform spectrograms in two SNR bins and, in addition, show the corresponding time-domain samples to cross-check amplitude and phase consistency (Fig.~\ref{fig:ts_samples}).

\begin{figure}[htbp]
  \centering
  \subfloat[Q-transform (SNR 1--5). The QNM ridge at \(f\simeq0.08\!-\!0.12\,\mathrm{Hz}\) is disrupted by the gap windows (vertical dotted lines). The noisy input shows a high-frequency noise spectrum (\(f\gtrsim0.3\,\mathrm{Hz}\)) and a low-frequency comb below \(3\times10^{-2}\,\mathrm{Hz}\). DGF restores ridge continuity and suppresses both artifacts (colorbar \(\times10^{-46}\)).\label{fig:spec_q_1-5}]{
    \includegraphics[width=1\linewidth]{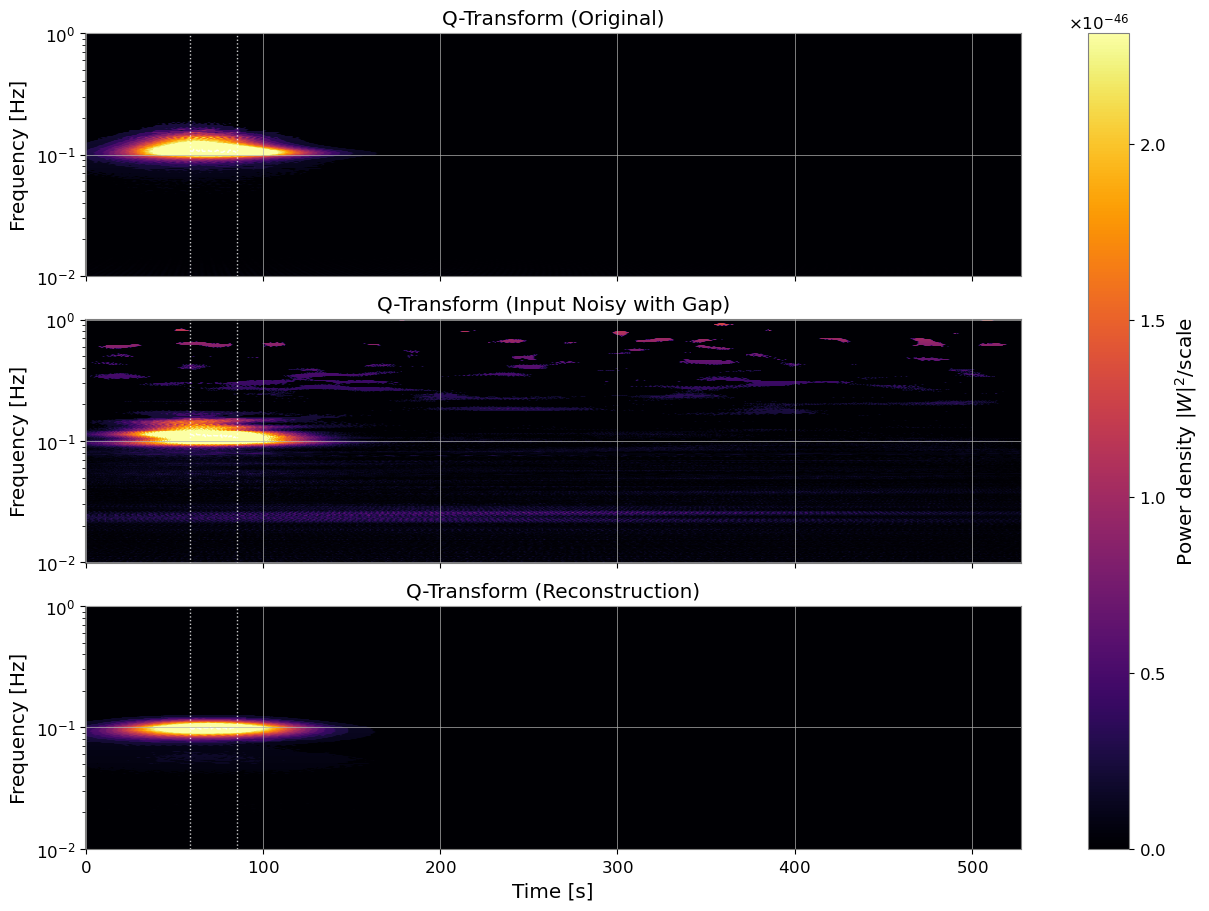}
  }\hfill
  \subfloat[Q-transform (SNR 5--10). The ridge sits at \(f\simeq(2\!-\!3)\times10^{-2}\,\mathrm{Hz}\). The gapped input shows a faint band above the ridge and mild smearing near the gap; DGF recovers a narrow, uniform ridge and removes the above-ridge leakage (colorbar \(\times10^{-45}\)).\label{fig:spec_q_5-10}]{
    \includegraphics[width=1\linewidth]{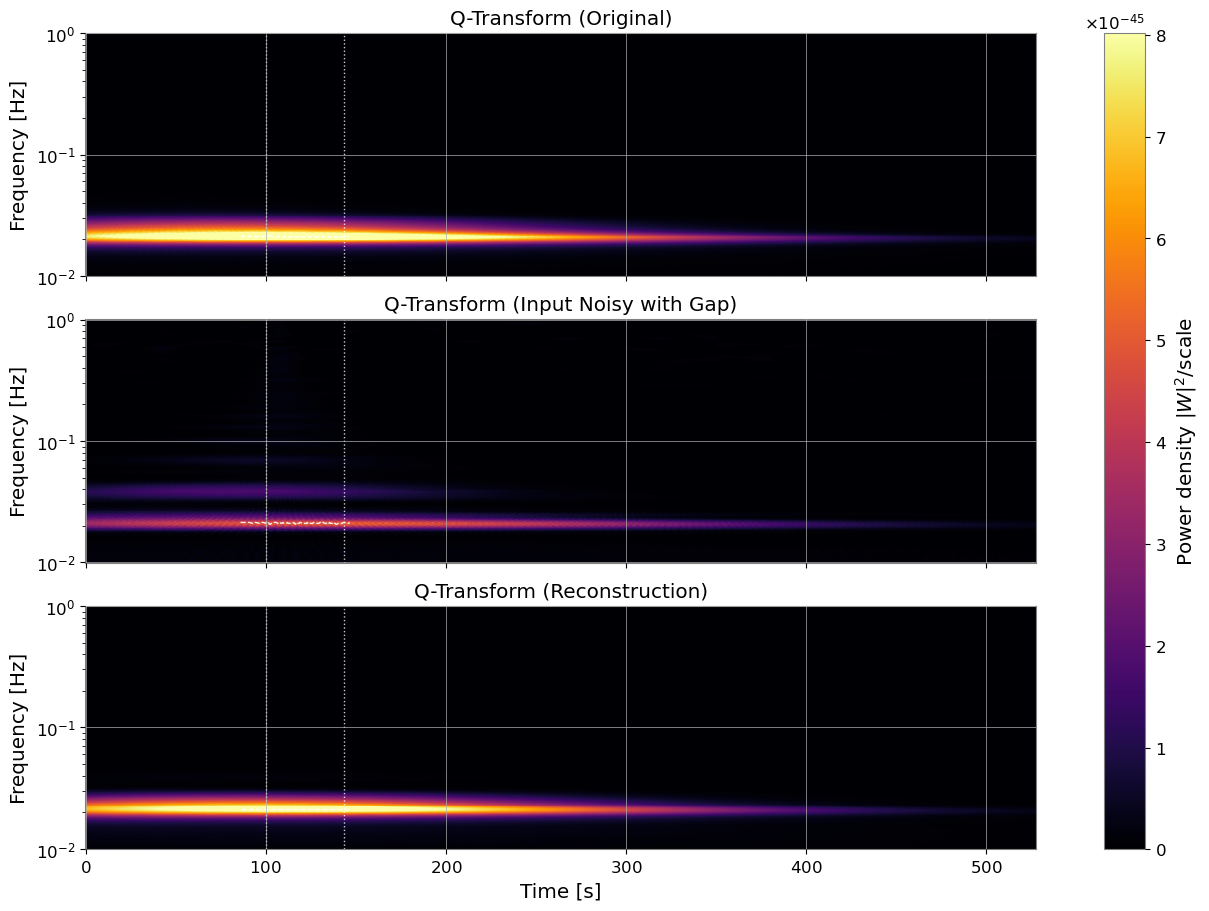}
  }
  \caption{\textbf{Q-transform spectrograms of the original signal (top), noisy gapped input (middle), and DGF reconstruction (bottom).}
  Across both SNR bins, DGF preserves the narrow QNM ridge, restores continuity across the gap windows, and suppresses spurious bands while maintaining time--frequency coherence.}
  \label{fig:spec_q}
\end{figure}

\begin{figure}[htbp]
  \centering
  \subfloat[Time-domain sample (SNR 1--5). The reconstructed waveform (green) recovers the burst-like ringdown near $t\!\sim\!60$--$120\,\mathrm{s}$, closely following the target (orange) while denoising the input (blue). Vertical scale $\times 10^{-22}$.%
  \label{fig:ts_snr15}]{
    \includegraphics[width=1\linewidth]{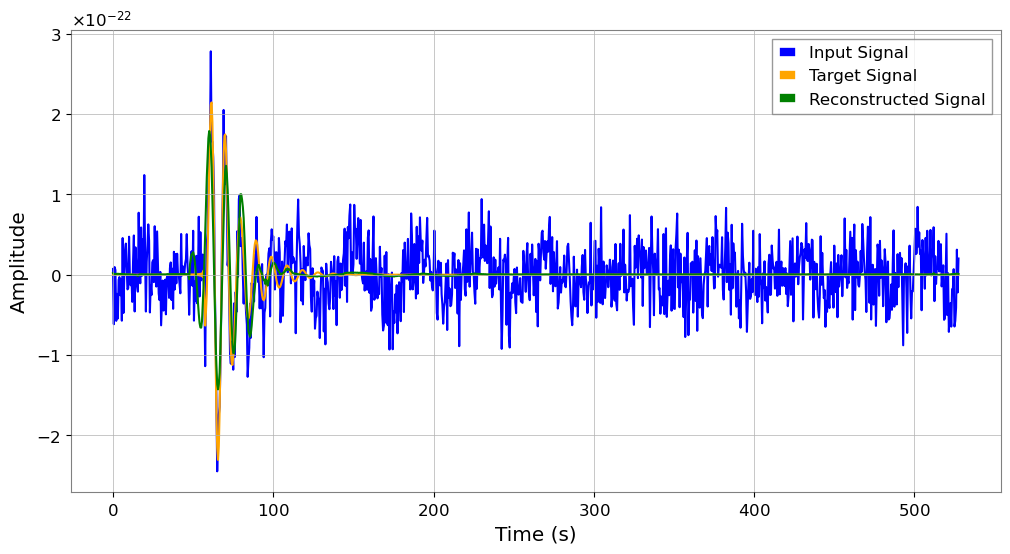}
  }\hfill
  \subfloat[Time-domain sample (SNR 5--10). The reconstruction tracks both amplitude and phase of multiple ringdown cycles and agrees with the target at the $\times 10^{-21}$ scale.%
  \label{fig:ts_snr510}]{
    \includegraphics[width=1\linewidth]{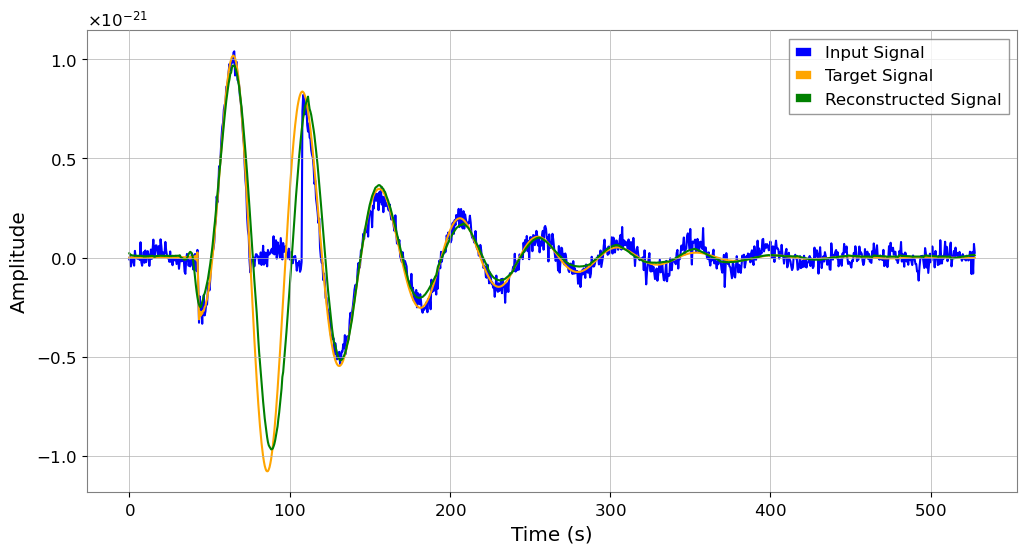}
  }
  \caption{\textbf{Time-domain samples corresponding to the spectrograms in Fig.~\ref{fig:spec_q}.}
  Panels (a) and (b) align with the SNR\,1--5 and SNR\,5--10 cases, respectively.}
  \label{fig:ts_samples}
\end{figure}

Figure~\ref{fig:spec_q_1-5} shows the lower-SNR case. From top to bottom are the original signal, the noisy gapped input, and the DGF reconstruction. 
The QNM ridge is concentrated at \(f\simeq(0.08\!-\!0.12)\,\mathrm{Hz}\) during \(t\sim40\)–\(140\,\mathrm{s}\), with the gap windows located around \(t\approx60\)–\(100\,\mathrm{s}\) (vertical dotted lines). 
In the gapped input (middle), two artifacts are evident: 
(i) a high-frequency noise spectrum consisting of intermittent blobs and streaks above the ridge in \(f\gtrsim0.3\,\mathrm{Hz}\), and 
(ii) a low-frequency comb-like background below \(3\times10^{-2}\,\mathrm{Hz}\) due to spectral leakage. 
Both effects disrupt the ridge continuity across the gaps. 
After reconstruction (bottom), the ridge becomes continuous through the gap windows and the above-ridge noise is strongly suppressed to near the colorbar floor (colorbar annotated in units of power density, \(\times10^{-46}\)). 
The associated time-domain sample in Fig.~\ref{fig:ts_samples}a (scale \(\times10^{-22}\)) shows that DGF recovers the ringdown burst amplitude and phase around \(t\!\sim\!60\)–\(120\,\mathrm{s}\) while removing surrounding noise fluctuations.

Figure~\ref{fig:spec_q_5-10} presents the higher-SNR case. 
Here the QNM ridge lies at a lower frequency, \(f\simeq(2\!-\!3)\times10^{-2}\,\mathrm{Hz}\), and is already clean in the original panel. 
The gapped input exhibits a faint, broad band above the ridge (\(f\gtrsim3\times10^{-2}\,\mathrm{Hz}\)) and mild ridge smearing near the gap window around \(t\approx110\)–\(170\,\mathrm{s}\). 
DGF restores a uniform, narrow ridge along time and suppresses the above-ridge leakage; the peak power along the ridge matches the original within the color-scale resolution (colorbar \(\times10^{-45}\)). 
Consistently, the time-domain sample in Fig.~\ref{fig:ts_samples}b (scale \(\times10^{-21}\)) shows that the reconstructed waveform tracks the target’s amplitude envelope and oscillation phase across multiple cycles.

Overall, across both SNR regimes the reconstructed ridge width, slope (time–frequency chirp), and peak power are visually consistent with the ground truth, indicating that DGF preserves time–frequency coherence while inpainting missing segments.

The spectral artifacts seen in the gapped input are the expected consequence of
multiplying the true waveform by a time window $W(t)$ that vanishes inside each gap:
$s_{\rm gap}(t)=W(t)h(t)+n(t)$. By the convolution theorem, the corresponding
Fourier-domain data obey $\tilde{s}_{\rm gap}(f)=\tilde{W}(f)*\tilde{h}(f)+\tilde{n}(f)$,
so the narrow QNM line (with width set by the damping time, {\rm FWHM} (Full Width at Half Maximum) $\approx\!1/(\pi\tau)$) is
broadened and its energy is re-distributed into side lobes---the familiar phenomenon of spectral
leakage in windowed Fourier analysis \cite{Harris:1978}. In a time--frequency representation with
approximately constant quality factor (e.g.\ the Q-transform used here), this manifests as a broken
or thickened ridge and spurious power above/below the physical band \cite{Chatterji:2004}.

To quantify the recovery of band-limited signal content, let $B$ denote a narrow
ringdown band around the ridge frequency $f_0$ with width $\Delta f\simeq {\rm FWHM}$, and define
the band energy (noise-weighted) and leakage fraction
\begin{equation}
\mathcal{E}_{B}(x)\equiv \int_{B}\frac{|\tilde{x}(f)|^{2}}{S_{n}(f)}\,df,\qquad
\Lambda(x)\equiv 1-\frac{\mathcal{E}_{B}(x)}{\int_{0}^{\infty}\frac{|\tilde{x}(f)|^{2}}{S_{n}(f)}\,df}.
\label{eq:band-metrics}
\end{equation}
For gap-free data with a single damped sinusoid, $\Lambda(h)$ is minimal and
$\mathcal{E}_{B}(h)$ captures essentially all of the mode power; introducing gaps increases
$\Lambda(s_{\rm gap})$ by convolving $\tilde{h}$ with $\tilde{W}$, thereby smearing the ridge.
After reconstruction, we observe both a reduction of $\Lambda(\hat h)$ and an increase of
$\mathcal{E}_{B}(\hat h)$ toward $\mathcal{E}_{B}(h)$, consistent with the visual restoration of a
uniform, narrow ridge and with the time-domain phase continuity. From a physical standpoint,
recovering the ringdown band restores the effective quality factor $Q\simeq \pi f_0\tau$ and the
mode’s peak power, which in turn improves mode separability and stabilizes joint inference of
$(\omega_R,\tau,A_{\ell mn})$ in ringdown spectroscopy \cite{Berti:2009}.
The agreement between the reconstructed and original ridges across both SNR bins, together with
the band-power recovery in Eq.~\eqref{eq:band-metrics}, indicates that the inpainting is not merely
interpolating missing samples but is effectively reversing the window-induced convolution that causes
spectral leakage, thereby preserving time--frequency coherence in the physically relevant QNM band.

\subsection{Phase Recovery and Alignment}\label{subsec:PhaseRecoveryAlignment}
Phase coherence is a critical feature of ringdown signals, as the extraction of QNM frequencies and damping times often relies on accurate instantaneous phase estimation \cite{135376}. 

\begin{figure}[htbp]
    \centering
    \includegraphics[width=0.5\textwidth]{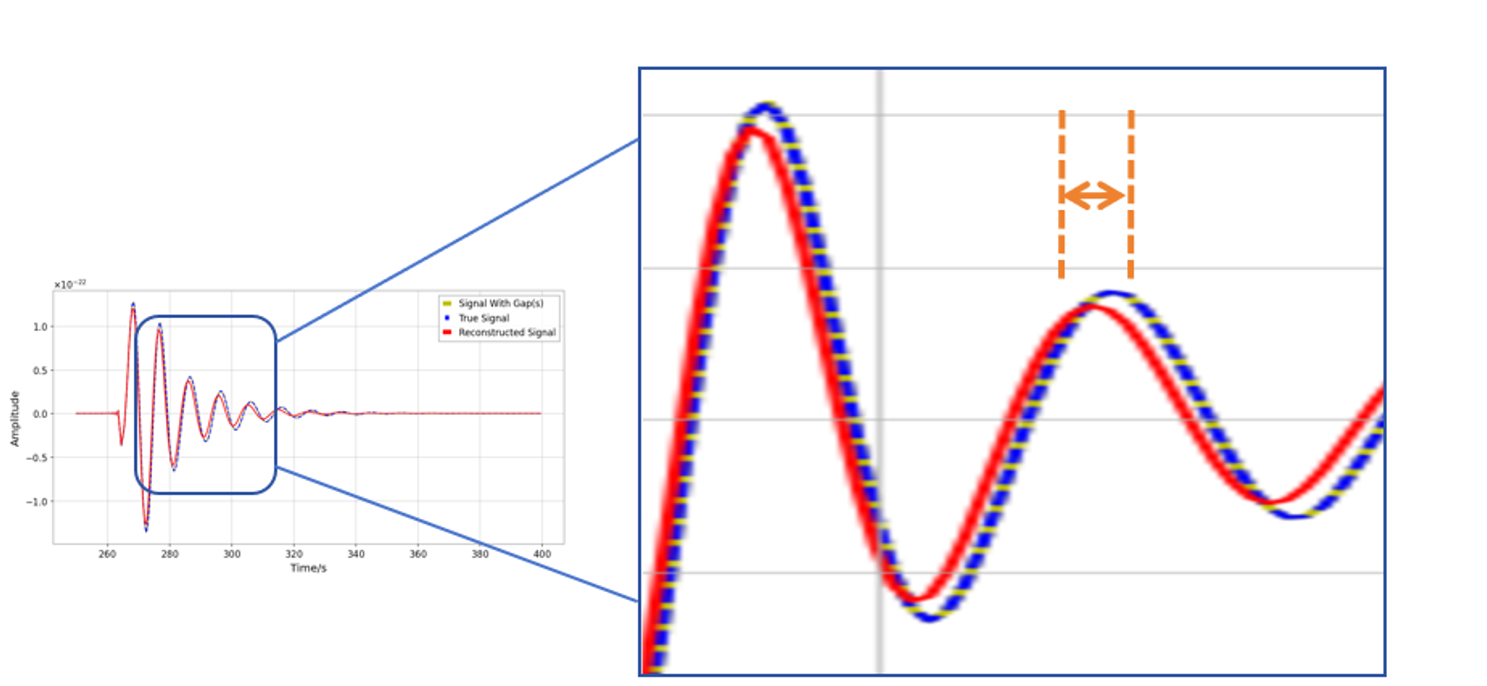}
    \caption{\textbf{Phase Recovery}: Detailed view of the ringdown peak shows the true waveform (blue dashed) and the reconstructed signal (red).  The orange arrows indicate the residual phase offset $\Delta\phi$ between the true and reconstructed peaks.}
    \label{fig:phase_recovery}
\end{figure}
A close-up on the fitted ringdown peak demonstrates that our model not only recovers the amplitude envelope but also aligns the waveform phase $\Delta\phi$ precisely. By overlaying the true and reconstructed signals, we see that the reconstructed crest (red) tracks the true crest (blue) with minimal lag, highlighting the network’s ability to reproduce the oscillatory phase structure critical for accurate parameter estimation, shown in Fig.~\ref{fig:phase_recovery}.

We obtain the analytic signal via the Hilbert transform,
\[
\mathcal{A}(t) = h(t) + i\,\mathcal{H}[h(t)],\quad
\phi(t) = \arg\bigl(\mathcal{A}(t)\bigr),
\]
where \(\mathcal{H}[\cdot]\) denotes the Hilbert transform.

For each test sample we compute:
\begin{enumerate}
  \item The absolute phase deviation at the peak amplitude \(t_p\):
  \[
    \Delta\phi_{\rm peak} = \bigl|\phi_{\rm recon}(t_p) - \phi_{\rm true}(t_p)\bigr|;
  \]
  \item The mean absolute phase deviation over the full sequence:
  \[
    \Delta\phi_{\rm mean} = \frac{1}{N}\sum_{n}\bigl|\phi_{\rm recon}(t_n) - \phi_{\rm true}(t_n)\bigr|,
  \]
\end{enumerate}
where $N$ represents the number of the top ($\left\lvert\mathcal{A}\right\lvert$) secondary peaks of signal under consideration. Here we set $N=5$. 

\begin{figure}[htbp]
  \centering
  \subfloat[Peak-phase deviation : mean $=-0.47^\circ$.%
  \label{fig:phase_peak1-5}]{
    \includegraphics[width=0.48\linewidth]{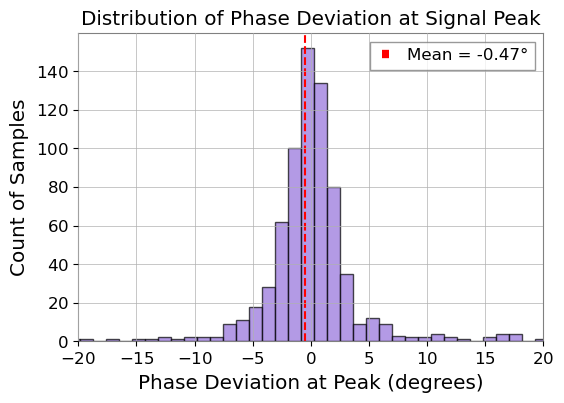}
  }\hfill
  \subfloat[Secondary peak absolute phase deviation: mean $=3.71^\circ$.%
  \label{fig:phase_mean_ab1-5}]{
    \includegraphics[width=0.48\linewidth]{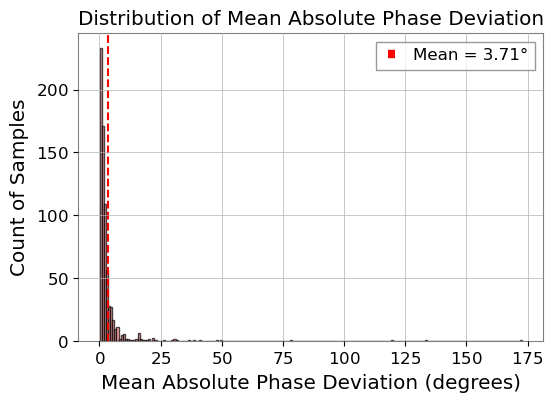}
  }
  \caption{\textbf{Histograms of phase deviation at the signal peaks for lower-SNR (1--5) group.}}
  \label{fig:phase_1-5}
\end{figure}

\begin{figure}[htbp]
  \centering
  \subfloat[Peak-phase deviation : mean $=-0.01^\circ$.%
  \label{fig:phase_peak5-10}]{
    \includegraphics[width=0.48\linewidth]{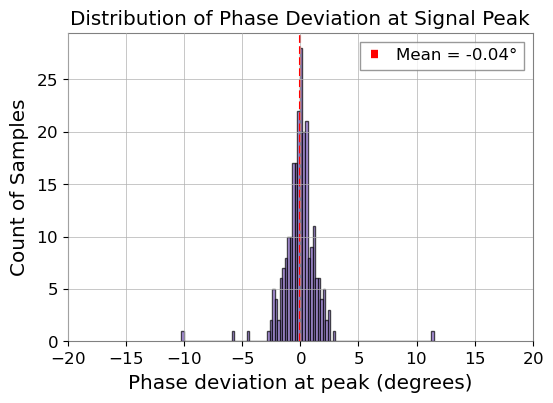}
  }\hfill
  \subfloat[Secondary peak absolute phase deviation: mean $=0.90^\circ$.%
  \label{fig:phase_mean_ab5-10}]{
    \includegraphics[width=0.48\linewidth]{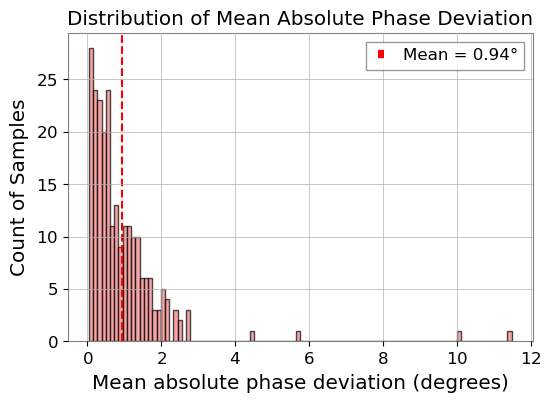}
  }
  \caption{\textbf{Histograms of phase deviation at the signal peaks for higher-SNR (5--10) group.}}
  \label{fig:phase_5-10}
\end{figure}

Fig.~\ref{fig:phase_1-5}, \ref{fig:phase_5-10} show histograms of the phase deviation \(\Delta\phi_{\rm peak}\) (in degrees) for the low and higher-SNR regimes. In the lower-SNR (1–5) subset (Fig.~\ref{fig:phase_1-5}), Fig.~\ref{fig:phase_peak1-5} shows that the mean peak-phase deviation is \(-0.47^\circ\), with over 95\% of samples within \(\pm10^\circ\) and over 90\% within \(\pm6.4^\circ\). For mean absolute phase deviation, which is shown in Fig.~\ref{fig:phase_mean_ab1-5}, we also get a mean value of \(3.71^\circ\), with over 95\% of samples within \(12^\circ\) and over 90\% within \(6.3^\circ\).

Fig.~\ref{fig:phase_5-10} shows histograms of the peak-phase deviation \(\Delta\phi_{\rm peak}\) at the maximum of the signal (Fig.~\ref{fig:phase_peak5-10}) and the distribution of the mean absolute phase difference (Fig.~\ref{fig:phase_mean_ab5-10}) for the SNR\,5–10 subset. In Fig.~\ref{fig:phase_peak5-10}, the mean peak-phase deviation is \(-0.01^\circ\), with over 95\% of samples lying within \(\pm2.3^\circ\) and over 90\% within \(\pm1.7^\circ\). In Fig.~\ref{fig:phase_mean_ab5-10}, the mean absolute phase deviation is \(0.90^\circ\), with 95\% of events below \(2.17^\circ\) and 90\% below \(1.6^\circ\). 

Overall, in the lower-SNR (1–5) regime,the DGF model still achieves a mean peak-phase deviation of \(-0.47^\circ\), with 90\% of samples contained within \(\pm6.4^\circ\) and 95\% within \(\pm10^\circ\), and a mean absolute phase deviation of \(3.71^\circ\) (90\% below \(6.3^\circ\), 95\% below \(12^\circ\)). This robust performance under challenging noise conditions highlights DGF’s advantage in lower-SNR reconstruction. In the higher-SNR (5–10) regime, these metrics tighten further to a mean peak-phase deviation of \(-0.01^\circ\) (90\% within \(\pm1.7^\circ\), 95\% within \(\pm2.3^\circ\)) and a mean absolute deviation of \(0.90^\circ\) (90\% below \(1.6^\circ\), 95\% below \(2.17^\circ\)), demonstrating near–ideal phase fidelity as the signal becomes stronger.


For a damped quasi-sinusoid $h(t)\!\sim\! A\,e^{-t/\tau}\cos(\omega_R t+\phi_0)$, the instantaneous phase
$\phi(t)$ governs both the frequency estimate $\omega_R\simeq d\phi/dt$ and the stability of multimode fits.
Let the residual phase error be $\varepsilon(t)\!\equiv\!\phi_{\rm recon}(t)-\phi_{\rm true}(t)$. In the small-error
regime, the noise-weighted overlap admits the approximation
\begin{equation}
\mathcal{O}\;\equiv\;1-\mathcal{M}\;\simeq\;1-\tfrac{1}{2}\,\langle \varepsilon^2\rangle_w,\qquad
\langle a\rangle_w\!=\!\frac{\int a(t)\,w(t)\,dt}{\int w(t)\,dt},
\label{eq:phase-overlap}
\end{equation}
so tighter phase alignment ($\langle\varepsilon^2\rangle_w\!\downarrow$) directly maps into higher overlap and hence larger
effective SNR via $\rho_{\rm rec}\!\simeq\!\mathcal{O}\rho_{\rm opt}$ (cf.\ Sec.~\ref{subsec:Mismatch}). Because $\omega_R\!=\!d\phi/dt$,
a slowly varying phase bias induces a frequency bias $\delta\omega_R\simeq\langle d\varepsilon/dt\rangle_w$, while random
phase fluctuations set a variance that, under standard regularity conditions, scales as
\begin{equation}
{\rm Var}(\widehat{\omega}_R)\;\propto\;\frac{1}{\rho_{\rm eff}^{\,2}\,T_{\rm eff}^{\,2}},\qquad
{\rm Var}(\widehat{\tau})\;\propto\;\frac{1}{\rho_{\rm eff}^{\,2}\,T_{\rm eff}},
\label{eq:crlb-scaling}
\end{equation}
in accordance with Cram\'er--Rao/Bayesian bounds for single-(quasi)sine and exponentially damped sinusoids
\cite{Bretthorst1988}. Here $T_{\rm eff}$ is the effective
coherent baseline over the ringdown window and $\rho_{\rm eff}$ the in-band SNR; both increase when gaps are bridged with
phase continuity, shrinking uncertainties on $(\omega_R,\tau)$ and stabilizing joint fits for $A_{\ell mn}$. Moreover,
phase recovery governs coherent combination across segments/modes: if the residual jitter has rms $\sigma_\phi$, the
coherent gain carries the familiar phasor factor $e^{-\sigma_\phi^2/2}$, so the
sub-degree peak-phase errors observed here preserve near-optimal coherent SNR and enhance the resolvability of
secondary/overtones in ringdown spectroscopy \cite{Berti:2009}.

\section{Conclusion and Discussion}
\label{sec:discussion}

The results presented in Sec.~\ref{sec:results} demonstrate that the DGF model not only fills data gaps but also denoises and faithfully preserves the key physical features of black-hole ringdown waveforms across a broad range of SNR conditions. In the lower-SNR (1–5) regime, DGF achieves a mean waveform mismatch of $0.023$ (median $0.0049$, 75th percentile $0.015$), while in the higher-SNR (5–10) regime these metrics improve to $0.00246$ (median $0.00097$, 75th percentile $0.00139$). Time-domain residual amplitudes shrink from $\sigma \approx 1.12 \times 10^{-22}$ in the noisy input to $\sigma \approx 7.0 \times 10^{-23}$ after reconstruction, and the power spectral density shows that the broadband noise floor in the $0.01\text{--}1\,\mathrm{Hz}$ QNM band is suppressed by roughly one to two orders of magnitude. Time–frequency spectrograms corroborate these findings: across both SNR bins, DGF restores the narrow QNM ridge and preserves its chirp. In the lower-SNR (1–5) case it reconnects the ridge across the gap windows while suppressing a high-frequency spurious band and a low-frequency comb; in the higher-SNR (5–10) case it removes residual above-ridge leakage and yields ridge power and width visually indistinguishable from the ground truth within the color-scale resolution. Phase-error analysis is equally strong: in the lower-SNR subset the mean peak-phase deviation is $-0.47^\circ$ ($\approx 0.0082$\,rad), with $90\%$ of samples within $\pm 6.4^\circ$ and $95\%$ within $\pm 10^\circ$, and the mean absolute phase deviation is $3.71^\circ$ ($\approx 0.065$\,rad). In the higher-SNR subset the mean peak-phase deviation tightens to $-0.01^\circ$ ($\approx 0.00017$\,rad), with $90\%$ of samples within $\pm 1.7^\circ$ and $95\%$ within $\pm 2.3^\circ$, and the mean absolute phase deviation falls to $0.90^\circ$ ($\approx 0.016$\,rad).

Together, these quantitative and qualitative findings underscore the power of a convolutional time-mixing architecture to infer damped sinusoidal structures directly from limited, corrupted waveforms—without any explicit spectral or phase‐guided loss. In particular, the model’s robust performance under challenging lower-SNR conditions highlights its promise as a preprocessing module for ringdown‐based tests of general relativity with both ground‐ and space‐based gravitational‐wave detectors.

\paragraph*{\textbf{Applicability and limitations.}}
DGF is most effective when the analyzed segment is ringdown–dominated and well approximated by a superposition of damped sinusoids, e.g., the post–merger phase of binary black hole coalescences in ground- and space-based data. Within this regime, it remains robust for gaps up to $\sim20\%$ of the ringdown duration and for network SNR as low as 1, reaching sub-percent median mismatches (median $\mathcal{M}\!\approx\!0.0049$ for SNR 1–5) and mean peak-phase deviations $<0.11$\,rad. Its lightweight convolutional time-mixing design enables low-latency inference and straightforward fine-tuning on QNM libraries spanning mass, spin, with potential extensions to multi-mode coupling and multi-channel denoising. The main caveats are: training currently relies on synthetic data with Gaussian noise, whereas real detectors can exhibit colored/nonstationary noise and glitches \cite{edwards2020nonstationary,baghi2019gaps}; the training parameter space is still limited in dimensionality and range, constraining generalization; slowly varying inspiral content and non-QNM transients are out of scope; and gap-coupled random phase noise is not explicitly modeled.

\paragraph*{\textbf{Comparison to Prior Work.}}
Unlike Xu \textit{et al.}~\cite{xu2024dense} and Mao \textit{et al.}~\cite{mao2024bigru},  DGF is optimized solely for the noisy ringdown data with gap, which demand a capability to handle dual tasks, but we can still obtain some encouraging results from comparison, such as mean overlap 0.996 in Ref\cite{xu2024dense} and over 0.99 in Ref\cite{mao2024bigru} while mismatches of DGF outputs come to a mean of 0.002 corresponding to a mean overlap of 0.998. This specialization allows a lean convolutional time-mixing architecture with significantly fewer parameters and lower inference latency, yet yields equal or better fidelity within the ringdown window.  In our tests, DGF achieves a mean mismatch of $\sim0.002$ and peak-phase deviation below $-0.01^\circ$ at SNR 5–10.


\paragraph*{\textbf{Future Work.}}

Several extensions are possible. First, integrating real instrument noise into the training set may improve generalization. Second, introducing parameterized QNM priors or physically informed losses could guide reconstruction further. Third, combining DGF with global IMR recovery modules could enable full waveform stitching across data gaps. What's more, The low-frequency characteristics of the gravitational waves we focus on will cause overlap among events. Therefore, it is of a great benefit to take the aliasing between different signals into account in the training set, and it also contributes to the future global fit project. Bayesian uncertainty estimation would allow confidence-based reconstructions loop with condition input, adjusting model by input new conditions inferred from reconstruction of last loop and finally converging to a stable result as expected, and this function is currently under development. Future work may extend DGF to handle multi-mode coupling, precession-induced frequency modulation, or joint denoising across multiple detector channels, further broadening its applicability to gravitational wave data analysis, which may provide a  great progress on generalization performance.




\begin{acknowledgments}
The authors thank He Wang and Yiyang Guo for helpful discussions. This research is supported in part by the National Key R\&D Program of China, grant number 2020YFC2201300, and the National Natural Science Foundation of China, grant numbers 12035016, 12375058 and 12361141825.

\end{acknowledgments}
\label{sec:acknowledgement}

\counterwithin{equation}{subsection}
\renewcommand{\thesubsection}{b}

\newpage
\bibliography{DGF_references}

\begin{thebibliography}{64}%
\makeatletter
\providecommand \@ifxundefined [1]{%
 \@ifx{#1\undefined}
}%
\providecommand \@ifnum [1]{%
 \ifnum #1\expandafter \@firstoftwo
 \else \expandafter \@secondoftwo
 \fi
}%
\providecommand \@ifx [1]{%
 \ifx #1\expandafter \@firstoftwo
 \else \expandafter \@secondoftwo
 \fi
}%
\providecommand \natexlab [1]{#1}%
\providecommand \enquote  [1]{``#1''}%
\providecommand \bibnamefont  [1]{#1}%
\providecommand \bibfnamefont [1]{#1}%
\providecommand \citenamefont [1]{#1}%
\providecommand \href@noop [0]{\@secondoftwo}%
\providecommand \href [0]{\begingroup \@sanitize@url \@href}%
\providecommand \@href[1]{\@@startlink{#1}\@@href}%
\providecommand \@@href[1]{\endgroup#1\@@endlink}%
\providecommand \@sanitize@url [0]{\catcode `\\12\catcode `\$12\catcode `\&12\catcode `\#12\catcode `\^12\catcode `\_12\catcode `\%12\relax}%
\providecommand \@@startlink[1]{}%
\providecommand \@@endlink[0]{}%
\providecommand \url  [0]{\begingroup\@sanitize@url \@url }%
\providecommand \@url [1]{\endgroup\@href {#1}{\urlprefix }}%
\providecommand \urlprefix  [0]{URL }%
\providecommand \Eprint [0]{\href }%
\providecommand \doibase [0]{https://doi.org/}%
\providecommand \selectlanguage [0]{\@gobble}%
\providecommand \bibinfo  [0]{\@secondoftwo}%
\providecommand \bibfield  [0]{\@secondoftwo}%
\providecommand \translation [1]{[#1]}%
\providecommand \BibitemOpen [0]{}%
\providecommand \bibitemStop [0]{}%
\providecommand \bibitemNoStop [0]{.\EOS\space}%
\providecommand \EOS [0]{\spacefactor3000\relax}%
\providecommand \BibitemShut  [1]{\csname bibitem#1\endcsname}%
\let\auto@bib@innerbib\@empty
\bibitem [{\citenamefont {Abbott}\ \emph {et~al.}(2016)\citenamefont {Abbott}, \citenamefont {Abbott}, \citenamefont {Abbott}, \citenamefont {Abernathy}, \citenamefont {Acernese}, \citenamefont {Ackley}, \citenamefont {Adams}, \citenamefont {Adams}, \citenamefont {Addesso}, \citenamefont {Adhikari} \emph {et~al.}}]{abbott2016observation}%
  \BibitemOpen
  \bibfield  {author} {\bibinfo {author} {\bibfnamefont {B.~P.}\ \bibnamefont {Abbott}}, \bibinfo {author} {\bibfnamefont {R.}~\bibnamefont {Abbott}}, \bibinfo {author} {\bibfnamefont {T.~D.}\ \bibnamefont {Abbott}}, \bibinfo {author} {\bibfnamefont {M.~R.}\ \bibnamefont {Abernathy}}, \bibinfo {author} {\bibfnamefont {F.}~\bibnamefont {Acernese}}, \bibinfo {author} {\bibfnamefont {K.}~\bibnamefont {Ackley}}, \bibinfo {author} {\bibfnamefont {C.}~\bibnamefont {Adams}}, \bibinfo {author} {\bibfnamefont {T.}~\bibnamefont {Adams}}, \bibinfo {author} {\bibfnamefont {P.}~\bibnamefont {Addesso}}, \bibinfo {author} {\bibfnamefont {R.~X.}\ \bibnamefont {Adhikari}}, \emph {et~al.},\ }\bibfield  {title} {\bibinfo {title} {Observation of gravitational waves from a binary black hole merger},\ }\href@noop {} {\bibfield  {journal} {\bibinfo  {journal} {Physical review letters}\ }\textbf {\bibinfo {volume} {116}},\ \bibinfo {pages} {061102} (\bibinfo {year} {2016})}\BibitemShut {NoStop}%
\bibitem [{\citenamefont {Amaro-Seoane}\ \emph {et~al.}(2017)\citenamefont {Amaro-Seoane}, \citenamefont {Audley}, \citenamefont {Babak}, \citenamefont {Baker}, \citenamefont {Barausse}, \citenamefont {Bender}, \citenamefont {Berti}, \citenamefont {Binetruy}, \citenamefont {Born}, \citenamefont {Bortoluzzi}, \citenamefont {Camp}, \citenamefont {Caprini}, \citenamefont {Cardoso}, \citenamefont {Colpi}, \citenamefont {Conklin}, \citenamefont {Cornish}, \citenamefont {Cutler}, \citenamefont {Danzmann}, \citenamefont {Dolesi}, \citenamefont {Ferraioli}, \citenamefont {Ferroni}, \citenamefont {Fitzsimons}, \citenamefont {Gair}, \citenamefont {Bote}, \citenamefont {Giardini}, \citenamefont {Gibert}, \citenamefont {Grimani}, \citenamefont {Halloin}, \citenamefont {Heinzel}, \citenamefont {Hertog}, \citenamefont {Hewitson}, \citenamefont {Holley-Bockelmann}, \citenamefont {Hollington}, \citenamefont {Hueller}, \citenamefont {Inchauspe}, \citenamefont {Jetzer}, \citenamefont {Karnesis}, \citenamefont {Killow},
  \citenamefont {Klein}, \citenamefont {Klipstein}, \citenamefont {Korsakova}, \citenamefont {Larson}, \citenamefont {Livas}, \citenamefont {Lloro}, \citenamefont {Man}, \citenamefont {Mance}, \citenamefont {Martino}, \citenamefont {Mateos}, \citenamefont {McKenzie}, \citenamefont {McWilliams}, \citenamefont {Miller}, \citenamefont {Mueller}, \citenamefont {Nardini}, \citenamefont {Nelemans}, \citenamefont {Nofrarias}, \citenamefont {Petiteau}, \citenamefont {Pivato}, \citenamefont {Plagnol}, \citenamefont {Porter}, \citenamefont {Reiche}, \citenamefont {Robertson}, \citenamefont {Robertson}, \citenamefont {Rossi}, \citenamefont {Russano}, \citenamefont {Schutz}, \citenamefont {Sesana}, \citenamefont {Shoemaker}, \citenamefont {Slutsky}, \citenamefont {Sopuerta}, \citenamefont {Sumner}, \citenamefont {Tamanini}, \citenamefont {Thorpe}, \citenamefont {Troebs}, \citenamefont {Vallisneri}, \citenamefont {Vecchio}, \citenamefont {Vetrugno}, \citenamefont {Vitale}, \citenamefont {Volonteri}, \citenamefont
  {Wanner}, \citenamefont {Ward}, \citenamefont {Wass}, \citenamefont {Weber}, \citenamefont {Ziemer},\ and\ \citenamefont {Zweifel}}]{amaroseoane2017laserinterferometerspaceantenna}%
  \BibitemOpen
  \bibfield  {author} {\bibinfo {author} {\bibfnamefont {P.}~\bibnamefont {Amaro-Seoane}}, \bibinfo {author} {\bibfnamefont {H.}~\bibnamefont {Audley}}, \bibinfo {author} {\bibfnamefont {S.}~\bibnamefont {Babak}}, \bibinfo {author} {\bibfnamefont {J.}~\bibnamefont {Baker}}, \bibinfo {author} {\bibfnamefont {E.}~\bibnamefont {Barausse}}, \bibinfo {author} {\bibfnamefont {P.}~\bibnamefont {Bender}}, \bibinfo {author} {\bibfnamefont {E.}~\bibnamefont {Berti}}, \bibinfo {author} {\bibfnamefont {P.}~\bibnamefont {Binetruy}}, \bibinfo {author} {\bibfnamefont {M.}~\bibnamefont {Born}}, \bibinfo {author} {\bibfnamefont {D.}~\bibnamefont {Bortoluzzi}}, \bibinfo {author} {\bibfnamefont {J.}~\bibnamefont {Camp}}, \bibinfo {author} {\bibfnamefont {C.}~\bibnamefont {Caprini}}, \bibinfo {author} {\bibfnamefont {V.}~\bibnamefont {Cardoso}}, \bibinfo {author} {\bibfnamefont {M.}~\bibnamefont {Colpi}}, \bibinfo {author} {\bibfnamefont {J.}~\bibnamefont {Conklin}}, \bibinfo {author} {\bibfnamefont {N.}~\bibnamefont {Cornish}},
  \bibinfo {author} {\bibfnamefont {C.}~\bibnamefont {Cutler}}, \bibinfo {author} {\bibfnamefont {K.}~\bibnamefont {Danzmann}}, \bibinfo {author} {\bibfnamefont {R.}~\bibnamefont {Dolesi}}, \bibinfo {author} {\bibfnamefont {L.}~\bibnamefont {Ferraioli}}, \bibinfo {author} {\bibfnamefont {V.}~\bibnamefont {Ferroni}}, \bibinfo {author} {\bibfnamefont {E.}~\bibnamefont {Fitzsimons}}, \bibinfo {author} {\bibfnamefont {J.}~\bibnamefont {Gair}}, \bibinfo {author} {\bibfnamefont {L.~G.}\ \bibnamefont {Bote}}, \bibinfo {author} {\bibfnamefont {D.}~\bibnamefont {Giardini}}, \bibinfo {author} {\bibfnamefont {F.}~\bibnamefont {Gibert}}, \bibinfo {author} {\bibfnamefont {C.}~\bibnamefont {Grimani}}, \bibinfo {author} {\bibfnamefont {H.}~\bibnamefont {Halloin}}, \bibinfo {author} {\bibfnamefont {G.}~\bibnamefont {Heinzel}}, \bibinfo {author} {\bibfnamefont {T.}~\bibnamefont {Hertog}}, \bibinfo {author} {\bibfnamefont {M.}~\bibnamefont {Hewitson}}, \bibinfo {author} {\bibfnamefont {K.}~\bibnamefont {Holley-Bockelmann}},
  \bibinfo {author} {\bibfnamefont {D.}~\bibnamefont {Hollington}}, \bibinfo {author} {\bibfnamefont {M.}~\bibnamefont {Hueller}}, \bibinfo {author} {\bibfnamefont {H.}~\bibnamefont {Inchauspe}}, \bibinfo {author} {\bibfnamefont {P.}~\bibnamefont {Jetzer}}, \bibinfo {author} {\bibfnamefont {N.}~\bibnamefont {Karnesis}}, \bibinfo {author} {\bibfnamefont {C.}~\bibnamefont {Killow}}, \bibinfo {author} {\bibfnamefont {A.}~\bibnamefont {Klein}}, \bibinfo {author} {\bibfnamefont {B.}~\bibnamefont {Klipstein}}, \bibinfo {author} {\bibfnamefont {N.}~\bibnamefont {Korsakova}}, \bibinfo {author} {\bibfnamefont {S.~L.}\ \bibnamefont {Larson}}, \bibinfo {author} {\bibfnamefont {J.}~\bibnamefont {Livas}}, \bibinfo {author} {\bibfnamefont {I.}~\bibnamefont {Lloro}}, \bibinfo {author} {\bibfnamefont {N.}~\bibnamefont {Man}}, \bibinfo {author} {\bibfnamefont {D.}~\bibnamefont {Mance}}, \bibinfo {author} {\bibfnamefont {J.}~\bibnamefont {Martino}}, \bibinfo {author} {\bibfnamefont {I.}~\bibnamefont {Mateos}}, \bibinfo
  {author} {\bibfnamefont {K.}~\bibnamefont {McKenzie}}, \bibinfo {author} {\bibfnamefont {S.~T.}\ \bibnamefont {McWilliams}}, \bibinfo {author} {\bibfnamefont {C.}~\bibnamefont {Miller}}, \bibinfo {author} {\bibfnamefont {G.}~\bibnamefont {Mueller}}, \bibinfo {author} {\bibfnamefont {G.}~\bibnamefont {Nardini}}, \bibinfo {author} {\bibfnamefont {G.}~\bibnamefont {Nelemans}}, \bibinfo {author} {\bibfnamefont {M.}~\bibnamefont {Nofrarias}}, \bibinfo {author} {\bibfnamefont {A.}~\bibnamefont {Petiteau}}, \bibinfo {author} {\bibfnamefont {P.}~\bibnamefont {Pivato}}, \bibinfo {author} {\bibfnamefont {E.}~\bibnamefont {Plagnol}}, \bibinfo {author} {\bibfnamefont {E.}~\bibnamefont {Porter}}, \bibinfo {author} {\bibfnamefont {J.}~\bibnamefont {Reiche}}, \bibinfo {author} {\bibfnamefont {D.}~\bibnamefont {Robertson}}, \bibinfo {author} {\bibfnamefont {N.}~\bibnamefont {Robertson}}, \bibinfo {author} {\bibfnamefont {E.}~\bibnamefont {Rossi}}, \bibinfo {author} {\bibfnamefont {G.}~\bibnamefont {Russano}}, \bibinfo
  {author} {\bibfnamefont {B.}~\bibnamefont {Schutz}}, \bibinfo {author} {\bibfnamefont {A.}~\bibnamefont {Sesana}}, \bibinfo {author} {\bibfnamefont {D.}~\bibnamefont {Shoemaker}}, \bibinfo {author} {\bibfnamefont {J.}~\bibnamefont {Slutsky}}, \bibinfo {author} {\bibfnamefont {C.~F.}\ \bibnamefont {Sopuerta}}, \bibinfo {author} {\bibfnamefont {T.}~\bibnamefont {Sumner}}, \bibinfo {author} {\bibfnamefont {N.}~\bibnamefont {Tamanini}}, \bibinfo {author} {\bibfnamefont {I.}~\bibnamefont {Thorpe}}, \bibinfo {author} {\bibfnamefont {M.}~\bibnamefont {Troebs}}, \bibinfo {author} {\bibfnamefont {M.}~\bibnamefont {Vallisneri}}, \bibinfo {author} {\bibfnamefont {A.}~\bibnamefont {Vecchio}}, \bibinfo {author} {\bibfnamefont {D.}~\bibnamefont {Vetrugno}}, \bibinfo {author} {\bibfnamefont {S.}~\bibnamefont {Vitale}}, \bibinfo {author} {\bibfnamefont {M.}~\bibnamefont {Volonteri}}, \bibinfo {author} {\bibfnamefont {G.}~\bibnamefont {Wanner}}, \bibinfo {author} {\bibfnamefont {H.}~\bibnamefont {Ward}}, \bibinfo {author}
  {\bibfnamefont {P.}~\bibnamefont {Wass}}, \bibinfo {author} {\bibfnamefont {W.}~\bibnamefont {Weber}}, \bibinfo {author} {\bibfnamefont {J.}~\bibnamefont {Ziemer}},\ and\ \bibinfo {author} {\bibfnamefont {P.}~\bibnamefont {Zweifel}},\ }\href {https://arxiv.org/abs/1702.00786} {\bibinfo {title} {Laser interferometer space antenna}} (\bibinfo {year} {2017}),\ \Eprint {https://arxiv.org/abs/1702.00786} {arXiv:1702.00786 [astro-ph.IM]} \BibitemShut {NoStop}%
\bibitem [{\citenamefont {Robson}\ \emph {et~al.}(2019{\natexlab{a}})\citenamefont {Robson}, \citenamefont {Cornish},\ and\ \citenamefont {Liu}}]{robson2019construction}%
  \BibitemOpen
  \bibfield  {author} {\bibinfo {author} {\bibfnamefont {T.}~\bibnamefont {Robson}}, \bibinfo {author} {\bibfnamefont {N.~J.}\ \bibnamefont {Cornish}},\ and\ \bibinfo {author} {\bibfnamefont {C.}~\bibnamefont {Liu}},\ }\bibfield  {title} {\bibinfo {title} {The construction and use of lisa sensitivity curves},\ }\href@noop {} {\bibfield  {journal} {\bibinfo  {journal} {Classical and Quantum Gravity}\ }\textbf {\bibinfo {volume} {36}},\ \bibinfo {pages} {105011} (\bibinfo {year} {2019}{\natexlab{a}})}\BibitemShut {NoStop}%
\bibitem [{\citenamefont {Berti}\ \emph {et~al.}(2006{\natexlab{a}})\citenamefont {Berti}, \citenamefont {Cardoso},\ and\ \citenamefont {Will}}]{berti2006gravitational}%
  \BibitemOpen
  \bibfield  {author} {\bibinfo {author} {\bibfnamefont {E.}~\bibnamefont {Berti}}, \bibinfo {author} {\bibfnamefont {V.}~\bibnamefont {Cardoso}},\ and\ \bibinfo {author} {\bibfnamefont {C.~M.}\ \bibnamefont {Will}},\ }\bibfield  {title} {\bibinfo {title} {Gravitational-wave spectroscopy of massive black holes with the space interferometer lisa},\ }\href@noop {} {\bibfield  {journal} {\bibinfo  {journal} {Physical Review D}\ }\textbf {\bibinfo {volume} {73}},\ \bibinfo {pages} {064030} (\bibinfo {year} {2006}{\natexlab{a}})}\BibitemShut {NoStop}%
\bibitem [{\citenamefont {Berti}\ \emph {et~al.}(2009{\natexlab{a}})\citenamefont {Berti}, \citenamefont {Cardoso},\ and\ \citenamefont {Starinets}}]{berti2009qnm}%
  \BibitemOpen
  \bibfield  {author} {\bibinfo {author} {\bibfnamefont {E.}~\bibnamefont {Berti}}, \bibinfo {author} {\bibfnamefont {V.}~\bibnamefont {Cardoso}},\ and\ \bibinfo {author} {\bibfnamefont {A.~O.}\ \bibnamefont {Starinets}},\ }\bibfield  {title} {\bibinfo {title} {Quasinormal modes of black holes and black branes},\ }\href@noop {} {\bibfield  {journal} {\bibinfo  {journal} {Classical and Quantum Gravity}\ }\textbf {\bibinfo {volume} {26}},\ \bibinfo {pages} {163001} (\bibinfo {year} {2009}{\natexlab{a}})}\BibitemShut {NoStop}%
\bibitem [{\citenamefont {Isi}\ \emph {et~al.}(2019)\citenamefont {Isi}, \citenamefont {Giesler}, \citenamefont {Farr}, \citenamefont {Scheel},\ and\ \citenamefont {Teukolsky}}]{Isi_2019}%
  \BibitemOpen
  \bibfield  {author} {\bibinfo {author} {\bibfnamefont {M.}~\bibnamefont {Isi}}, \bibinfo {author} {\bibfnamefont {M.}~\bibnamefont {Giesler}}, \bibinfo {author} {\bibfnamefont {W.~M.}\ \bibnamefont {Farr}}, \bibinfo {author} {\bibfnamefont {M.~A.}\ \bibnamefont {Scheel}},\ and\ \bibinfo {author} {\bibfnamefont {S.~A.}\ \bibnamefont {Teukolsky}},\ }\bibfield  {title} {\bibinfo {title} {Testing the no-hair theorem with gw150914},\ }\bibfield  {journal} {\bibinfo  {journal} {Physical Review Letters}\ }\textbf {\bibinfo {volume} {123}},\ \href {https://doi.org/10.1103/physrevlett.123.111102} {10.1103/physrevlett.123.111102} (\bibinfo {year} {2019})\BibitemShut {NoStop}%
\bibitem [{\citenamefont {Baghi}\ \emph {et~al.}(2019{\natexlab{a}})\citenamefont {Baghi}, \citenamefont {Crowder},\ and\ \citenamefont {Vecchio}}]{baghi2019gaps}%
  \BibitemOpen
  \bibfield  {author} {\bibinfo {author} {\bibfnamefont {A.}~\bibnamefont {Baghi}}, \bibinfo {author} {\bibfnamefont {J.}~\bibnamefont {Crowder}},\ and\ \bibinfo {author} {\bibfnamefont {A.}~\bibnamefont {Vecchio}},\ }\bibfield  {title} {\bibinfo {title} {{Bayesian inference for non-stationary data gaps in LISA}},\ }\href {https://doi.org/10.1103/PhysRevD.100.046017} {\bibfield  {journal} {\bibinfo  {journal} {Phys.\ Rev.\ D}\ }\textbf {\bibinfo {volume} {100}},\ \bibinfo {pages} {046017} (\bibinfo {year} {2019}{\natexlab{a}})}\BibitemShut {NoStop}%
\bibitem [{\citenamefont {Dey}\ \emph {et~al.}(2021)\citenamefont {Dey}, \citenamefont {Karnesis}, \citenamefont {Toubiana}, \citenamefont {Barausse}, \citenamefont {Korsakova}, \citenamefont {Baghi},\ and\ \citenamefont {Basak}}]{dey2021gaps}%
  \BibitemOpen
  \bibfield  {author} {\bibinfo {author} {\bibfnamefont {K.}~\bibnamefont {Dey}}, \bibinfo {author} {\bibfnamefont {N.}~\bibnamefont {Karnesis}}, \bibinfo {author} {\bibfnamefont {A.}~\bibnamefont {Toubiana}}, \bibinfo {author} {\bibfnamefont {E.}~\bibnamefont {Barausse}}, \bibinfo {author} {\bibfnamefont {N.}~\bibnamefont {Korsakova}}, \bibinfo {author} {\bibfnamefont {Q.}~\bibnamefont {Baghi}},\ and\ \bibinfo {author} {\bibfnamefont {S.}~\bibnamefont {Basak}},\ }\bibfield  {title} {\bibinfo {title} {Effect of data gaps on the detectability and parameter estimation of massive black hole binaries with lisa},\ }\href@noop {} {\bibfield  {journal} {\bibinfo  {journal} {Physical Review D}\ }\textbf {\bibinfo {volume} {104}},\ \bibinfo {pages} {043004} (\bibinfo {year} {2021})}\BibitemShut {NoStop}%
\bibitem [{\citenamefont {Spadaro}\ \emph {et~al.}(2023)\citenamefont {Spadaro}, \citenamefont {Porter},\ and\ \citenamefont {Danzmann}}]{Spadaro:2023hlw}%
  \BibitemOpen
  \bibfield  {author} {\bibinfo {author} {\bibfnamefont {E.}~\bibnamefont {Spadaro}}, \bibinfo {author} {\bibfnamefont {E.~K.}\ \bibnamefont {Porter}},\ and\ \bibinfo {author} {\bibfnamefont {K.}~\bibnamefont {Danzmann}},\ }\bibfield  {title} {\bibinfo {title} {{Joint Parameter Estimation in the Presence of Time-Varying Non-stationarities for Space-based Gravitational Wave Detectors}},\ }\href {https://doi.org/10.1103/PhysRevD.107.064022} {\bibfield  {journal} {\bibinfo  {journal} {Phys.\ Rev.\ D}\ }\textbf {\bibinfo {volume} {107}},\ \bibinfo {pages} {064022} (\bibinfo {year} {2023})}\BibitemShut {NoStop}%
\bibitem [{\citenamefont {Shi}\ \emph {et~al.}(2024)\citenamefont {Shi}, \citenamefont {Jiao}, \citenamefont {Lai}, \citenamefont {Li}, \citenamefont {Shao},\ and\ \citenamefont {Tian}}]{shi2024effectsdatagapsringdown}%
  \BibitemOpen
  \bibfield  {author} {\bibinfo {author} {\bibfnamefont {J.}~\bibnamefont {Shi}}, \bibinfo {author} {\bibfnamefont {J.}~\bibnamefont {Jiao}}, \bibinfo {author} {\bibfnamefont {J.}~\bibnamefont {Lai}}, \bibinfo {author} {\bibfnamefont {Z.}~\bibnamefont {Li}}, \bibinfo {author} {\bibfnamefont {C.}~\bibnamefont {Shao}},\ and\ \bibinfo {author} {\bibfnamefont {Y.}~\bibnamefont {Tian}},\ }\href {https://arxiv.org/abs/2411.05415} {\bibinfo {title} {The effects of data gaps on ringdown signals with space-based joint observation}} (\bibinfo {year} {2024}),\ \Eprint {https://arxiv.org/abs/2411.05415} {arXiv:2411.05415 [gr-qc]} \BibitemShut {NoStop}%
\bibitem [{\citenamefont {Edwards}\ \emph {et~al.}(2020)\citenamefont {Edwards}, \citenamefont {Maturana-Russel}, \citenamefont {Meyer}, \citenamefont {Gair}, \citenamefont {Korsakova},\ and\ \citenamefont {Christensen}}]{edwards2020nonstationary}%
  \BibitemOpen
  \bibfield  {author} {\bibinfo {author} {\bibfnamefont {M.~C.}\ \bibnamefont {Edwards}}, \bibinfo {author} {\bibfnamefont {P.}~\bibnamefont {Maturana-Russel}}, \bibinfo {author} {\bibfnamefont {R.}~\bibnamefont {Meyer}}, \bibinfo {author} {\bibfnamefont {J.}~\bibnamefont {Gair}}, \bibinfo {author} {\bibfnamefont {N.}~\bibnamefont {Korsakova}},\ and\ \bibinfo {author} {\bibfnamefont {N.}~\bibnamefont {Christensen}},\ }\bibfield  {title} {\bibinfo {title} {Identifying and addressing nonstationary lisa noise},\ }\href@noop {} {\bibfield  {journal} {\bibinfo  {journal} {Physical Review D}\ }\textbf {\bibinfo {volume} {102}},\ \bibinfo {pages} {084062} (\bibinfo {year} {2020})}\BibitemShut {NoStop}%
\bibitem [{\citenamefont {Nissanke}\ \emph {et~al.}(2010)\citenamefont {Nissanke}, \citenamefont {Vallisneri}, \citenamefont {Kanner}, \citenamefont {Reitze},\ and\ \citenamefont {Dalya}}]{Nissanke:2010tm}%
  \BibitemOpen
  \bibfield  {author} {\bibinfo {author} {\bibfnamefont {S.}~\bibnamefont {Nissanke}}, \bibinfo {author} {\bibfnamefont {M.}~\bibnamefont {Vallisneri}}, \bibinfo {author} {\bibfnamefont {J.~A.}\ \bibnamefont {Kanner}}, \bibinfo {author} {\bibfnamefont {C.~M.}\ \bibnamefont {Reitze}},\ and\ \bibinfo {author} {\bibfnamefont {N.}~\bibnamefont {Dalya}},\ }\bibfield  {title} {\bibinfo {title} {{Exploring Short-Lived Binary Neutron Star Coalescences with a Network of Gravitational-Wave Detectors}},\ }\href@noop {} {\bibfield  {journal} {\bibinfo  {journal} {Astrophys.\ J.}\ }\textbf {\bibinfo {volume} {725}},\ \bibinfo {pages} {496–514} (\bibinfo {year} {2010})}\BibitemShut {NoStop}%
\bibitem [{\citenamefont {Zaldarriaga}\ and\ \citenamefont {Seljak}(2003)}]{Zaldarriaga:2003ip}%
  \BibitemOpen
  \bibfield  {author} {\bibinfo {author} {\bibfnamefont {M.}~\bibnamefont {Zaldarriaga}}\ and\ \bibinfo {author} {\bibfnamefont {U.}~\bibnamefont {Seljak}},\ }\bibfield  {title} {\bibinfo {title} {{An All-Sky Analysis of Standard Sirens with Gravitational-Wave Interferometers}},\ }\href {https://doi.org/10.1103/PhysRevD.67.043002} {\bibfield  {journal} {\bibinfo  {journal} {Phys.\ Rev.\ D}\ }\textbf {\bibinfo {volume} {67}},\ \bibinfo {pages} {043002} (\bibinfo {year} {2003})}\BibitemShut {NoStop}%
\bibitem [{\citenamefont {George}\ and\ \citenamefont {Huerta}(2018)}]{George:2017pmj}%
  \BibitemOpen
  \bibfield  {author} {\bibinfo {author} {\bibfnamefont {D.}~\bibnamefont {George}}\ and\ \bibinfo {author} {\bibfnamefont {E.~A.}\ \bibnamefont {Huerta}},\ }\bibfield  {title} {\bibinfo {title} {{Deep Learning for Real‐Time Gravitational Wave Detection and Parameter Estimation: Results with Advanced LIGO Data}},\ }\href {https://doi.org/10.1016/j.physletb.2017.11.070} {\bibfield  {journal} {\bibinfo  {journal} {Phys.\ Lett.\ B}\ }\textbf {\bibinfo {volume} {778}},\ \bibinfo {pages} {64–70} (\bibinfo {year} {2018})}\BibitemShut {NoStop}%
\bibitem [{\citenamefont {Horowitz}(1974)}]{Horowitz1974}%
  \BibitemOpen
  \bibfield  {author} {\bibinfo {author} {\bibfnamefont {L.~L.}\ \bibnamefont {Horowitz}},\ }\href {https://ntrs.nasa.gov/citations/19740041424} {\emph {\bibinfo {title} {The effects of spline interpolation on power spectral density}}},\ \bibinfo {type} {Tech. Rep.}\ (\bibinfo  {institution} {NASA},\ \bibinfo {year} {1974})\BibitemShut {NoStop}%
\bibitem [{\citenamefont {Maeland}(1988)}]{Maeland1988}%
  \BibitemOpen
  \bibfield  {author} {\bibinfo {author} {\bibfnamefont {E.}~\bibnamefont {Maeland}},\ }\bibfield  {title} {\bibinfo {title} {On the comparison of interpolation methods},\ }\href {https://doi.org/10.1109/42.730} {\bibfield  {journal} {\bibinfo  {journal} {IEEE Transactions on Medical Imaging}\ }\textbf {\bibinfo {volume} {7}},\ \bibinfo {pages} {213} (\bibinfo {year} {1988})}\BibitemShut {NoStop}%
\bibitem [{\citenamefont {Unser}(1999)}]{Unser1999}%
  \BibitemOpen
  \bibfield  {author} {\bibinfo {author} {\bibfnamefont {M.}~\bibnamefont {Unser}},\ }\bibfield  {title} {\bibinfo {title} {Splines—a perfect fit for signal and image processing},\ }\href {https://doi.org/10.1109/79.799930} {\bibfield  {journal} {\bibinfo  {journal} {IEEE Signal Processing Magazine}\ }\textbf {\bibinfo {volume} {16}},\ \bibinfo {pages} {22} (\bibinfo {year} {1999})}\BibitemShut {NoStop}%
\bibitem [{\citenamefont {Harris}(1978{\natexlab{a}})}]{Harris1978}%
  \BibitemOpen
  \bibfield  {author} {\bibinfo {author} {\bibfnamefont {F.~J.}\ \bibnamefont {Harris}},\ }\bibfield  {title} {\bibinfo {title} {On the use of windows for harmonic analysis with the discrete fourier transform},\ }\href {https://doi.org/10.1109/PROC.1978.10837} {\bibfield  {journal} {\bibinfo  {journal} {Proceedings of the IEEE}\ }\textbf {\bibinfo {volume} {66}},\ \bibinfo {pages} {51} (\bibinfo {year} {1978}{\natexlab{a}})}\BibitemShut {NoStop}%
\bibitem [{\citenamefont {Vallisneri}(2005)}]{Vallisneri2005}%
  \BibitemOpen
  \bibfield  {author} {\bibinfo {author} {\bibfnamefont {M.}~\bibnamefont {Vallisneri}},\ }\bibfield  {title} {\bibinfo {title} {Simulating time-delay interferometry in a model lisa},\ }\href {https://doi.org/10.1103/PhysRevD.71.022001} {\bibfield  {journal} {\bibinfo  {journal} {Physical Review D}\ }\textbf {\bibinfo {volume} {71}},\ \bibinfo {pages} {022001} (\bibinfo {year} {2005})}\BibitemShut {NoStop}%
\bibitem [{\citenamefont {Baghi}\ \emph {et~al.}(2019{\natexlab{b}})\citenamefont {Baghi}, \citenamefont {Thorpe}, \citenamefont {Slutsky}, \citenamefont {Baker}, \citenamefont {Canton}, \citenamefont {Korsakova},\ and\ \citenamefont {Karnesis}}]{Baghi2019}%
  \BibitemOpen
  \bibfield  {author} {\bibinfo {author} {\bibfnamefont {Q.}~\bibnamefont {Baghi}}, \bibinfo {author} {\bibfnamefont {J.~I.}\ \bibnamefont {Thorpe}}, \bibinfo {author} {\bibfnamefont {J.}~\bibnamefont {Slutsky}}, \bibinfo {author} {\bibfnamefont {J.}~\bibnamefont {Baker}}, \bibinfo {author} {\bibfnamefont {T.~D.}\ \bibnamefont {Canton}}, \bibinfo {author} {\bibfnamefont {N.}~\bibnamefont {Korsakova}},\ and\ \bibinfo {author} {\bibfnamefont {N.}~\bibnamefont {Karnesis}},\ }\bibfield  {title} {\bibinfo {title} {Gravitational-wave parameter estimation with gaps in lisa: A bayesian data augmentation method},\ }\href {https://doi.org/10.1103/PhysRevD.100.022003} {\bibfield  {journal} {\bibinfo  {journal} {Physical Review D}\ }\textbf {\bibinfo {volume} {100}},\ \bibinfo {pages} {022003} (\bibinfo {year} {2019}{\natexlab{b}})}\BibitemShut {NoStop}%
\bibitem [{\citenamefont {Blelly}\ \emph {et~al.}(2021)\citenamefont {Blelly}, \citenamefont {Bobin},\ and\ \citenamefont {Moutarde}}]{Blelly2021}%
  \BibitemOpen
  \bibfield  {author} {\bibinfo {author} {\bibfnamefont {A.}~\bibnamefont {Blelly}}, \bibinfo {author} {\bibfnamefont {J.}~\bibnamefont {Bobin}},\ and\ \bibinfo {author} {\bibfnamefont {H.}~\bibnamefont {Moutarde}},\ }\bibfield  {title} {\bibinfo {title} {Sparse data inpainting for the recovery of galactic-binary gravitational wave signals from gapped data},\ }\href {https://arxiv.org/abs/2104.05250} {\bibfield  {journal} {\bibinfo  {journal} {arXiv preprint}\ } (\bibinfo {year} {2021})},\ \Eprint {https://arxiv.org/abs/2104.05250} {2104.05250} \BibitemShut {NoStop}%
\bibitem [{\citenamefont {Starck}\ \emph {et~al.}(2010)\citenamefont {Starck}, \citenamefont {Murtagh},\ and\ \citenamefont {Fadili}}]{Starck2010}%
  \BibitemOpen
  \bibfield  {author} {\bibinfo {author} {\bibfnamefont {J.-L.}\ \bibnamefont {Starck}}, \bibinfo {author} {\bibfnamefont {F.}~\bibnamefont {Murtagh}},\ and\ \bibinfo {author} {\bibfnamefont {J.}~\bibnamefont {Fadili}},\ }\href {https://doi.org/10.1017/CBO9780511730344} {\emph {\bibinfo {title} {Sparse Image and Signal Processing: Wavelets, Curvelets, Morphological Diversity}}}\ (\bibinfo  {publisher} {Cambridge University Press},\ \bibinfo {year} {2010})\BibitemShut {NoStop}%
\bibitem [{\citenamefont {Xu}\ \emph {et~al.}(2024{\natexlab{a}})\citenamefont {Xu}, \citenamefont {Du}, \citenamefont {Xu}, \citenamefont {Liang},\ and\ \citenamefont {Wang}}]{Xu_2024}%
  \BibitemOpen
  \bibfield  {author} {\bibinfo {author} {\bibfnamefont {Y.}~\bibnamefont {Xu}}, \bibinfo {author} {\bibfnamefont {M.}~\bibnamefont {Du}}, \bibinfo {author} {\bibfnamefont {P.}~\bibnamefont {Xu}}, \bibinfo {author} {\bibfnamefont {B.}~\bibnamefont {Liang}},\ and\ \bibinfo {author} {\bibfnamefont {H.}~\bibnamefont {Wang}},\ }\bibfield  {title} {\bibinfo {title} {Gravitational wave signal extraction against non-stationary instrumental noises with deep neural network},\ }\href {https://doi.org/10.1016/j.physletb.2024.139016} {\bibfield  {journal} {\bibinfo  {journal} {Physics Letters B}\ }\textbf {\bibinfo {volume} {858}},\ \bibinfo {pages} {139016} (\bibinfo {year} {2024}{\natexlab{a}})}\BibitemShut {NoStop}%
\bibitem [{\citenamefont {Mao}\ \emph {et~al.}(2025)\citenamefont {Mao}, \citenamefont {Lee},\ and\ \citenamefont {Edwards}}]{Mao:2025cae}%
  \BibitemOpen
  \bibfield  {author} {\bibinfo {author} {\bibfnamefont {R.}~\bibnamefont {Mao}}, \bibinfo {author} {\bibfnamefont {J.~E.}\ \bibnamefont {Lee}},\ and\ \bibinfo {author} {\bibfnamefont {M.~C.}\ \bibnamefont {Edwards}},\ }\bibfield  {title} {\bibinfo {title} {{A novel stacked hybrid autoencoder for imputing LISA data gaps}},\ }\href@noop {} {\bibfield  {journal} {\bibinfo  {journal} {arXiv e-prints}\ } (\bibinfo {year} {2025})},\ \bibinfo {note} {v1},\ \Eprint {https://arxiv.org/abs/2410.05571} {arXiv:2410.05571 [gr-qc]} \BibitemShut {NoStop}%
\bibitem [{\citenamefont {Wang}\ \emph {et~al.}(2024)\citenamefont {Wang}, \citenamefont {Zhou}, \citenamefont {Cao}, \citenamefont {Guo},\ and\ \citenamefont {Ren}}]{Wang:2024waveformer}%
  \BibitemOpen
  \bibfield  {author} {\bibinfo {author} {\bibfnamefont {H.}~\bibnamefont {Wang}}, \bibinfo {author} {\bibfnamefont {Y.}~\bibnamefont {Zhou}}, \bibinfo {author} {\bibfnamefont {Z.}~\bibnamefont {Cao}}, \bibinfo {author} {\bibfnamefont {Z.-K.}\ \bibnamefont {Guo}},\ and\ \bibinfo {author} {\bibfnamefont {Z.}~\bibnamefont {Ren}},\ }\bibfield  {title} {\bibinfo {title} {{WaveFormer: transformer-based denoising method for gravitational-wave data}},\ }\href@noop {} {\bibfield  {journal} {\bibinfo  {journal} {arXiv e-prints}\ } (\bibinfo {year} {2024})},\ \bibinfo {note} {v2},\ \Eprint {https://arxiv.org/abs/2212.14283} {arXiv:2212.14283 [gr-qc]} \BibitemShut {NoStop}%
\bibitem [{\citenamefont {Chatterji}\ \emph {et~al.}(2004{\natexlab{a}})\citenamefont {Chatterji}, \citenamefont {Blackburn}, \citenamefont {Martin},\ and\ \citenamefont {Katsavounidis}}]{Chatterji:2004qtf}%
  \BibitemOpen
  \bibfield  {author} {\bibinfo {author} {\bibfnamefont {S.}~\bibnamefont {Chatterji}}, \bibinfo {author} {\bibfnamefont {L.}~\bibnamefont {Blackburn}}, \bibinfo {author} {\bibfnamefont {G.}~\bibnamefont {Martin}},\ and\ \bibinfo {author} {\bibfnamefont {E.}~\bibnamefont {Katsavounidis}},\ }\bibfield  {title} {\bibinfo {title} {{Multiresolution techniques for the detection of gravitational-wave bursts}},\ }\href {https://doi.org/10.1088/0264-9381/21/20/025} {\bibfield  {journal} {\bibinfo  {journal} {Class.\ Quant.\ Grav.}\ }\textbf {\bibinfo {volume} {21}},\ \bibinfo {pages} {S1809} (\bibinfo {year} {2004}{\natexlab{a}})}\BibitemShut {NoStop}%
\bibitem [{\citenamefont {Wang}\ \emph {et~al.}(2025)\citenamefont {Wang}, \citenamefont {Li}, \citenamefont {Shi}, \citenamefont {Ye}, \citenamefont {Mo}, \citenamefont {Lin}, \citenamefont {Ju}, \citenamefont {Chu},\ and\ \citenamefont {Jin}}]{wang2025timemixergeneraltimeseries}%
  \BibitemOpen
  \bibfield  {author} {\bibinfo {author} {\bibfnamefont {S.}~\bibnamefont {Wang}}, \bibinfo {author} {\bibfnamefont {J.}~\bibnamefont {Li}}, \bibinfo {author} {\bibfnamefont {X.}~\bibnamefont {Shi}}, \bibinfo {author} {\bibfnamefont {Z.}~\bibnamefont {Ye}}, \bibinfo {author} {\bibfnamefont {B.}~\bibnamefont {Mo}}, \bibinfo {author} {\bibfnamefont {W.}~\bibnamefont {Lin}}, \bibinfo {author} {\bibfnamefont {S.}~\bibnamefont {Ju}}, \bibinfo {author} {\bibfnamefont {Z.}~\bibnamefont {Chu}},\ and\ \bibinfo {author} {\bibfnamefont {M.}~\bibnamefont {Jin}},\ }\href {https://arxiv.org/abs/2410.16032} {\bibinfo {title} {Timemixer++: A general time series pattern machine for universal predictive analysis}} (\bibinfo {year} {2025}),\ \Eprint {https://arxiv.org/abs/2410.16032} {arXiv:2410.16032 [cs.LG]} \BibitemShut {NoStop}%
\bibitem [{\citenamefont {Devlin}\ \emph {et~al.}(2019)\citenamefont {Devlin}, \citenamefont {Chang}, \citenamefont {Lee},\ and\ \citenamefont {Toutanova}}]{devlin2019bertpretrainingdeepbidirectional}%
  \BibitemOpen
  \bibfield  {author} {\bibinfo {author} {\bibfnamefont {J.}~\bibnamefont {Devlin}}, \bibinfo {author} {\bibfnamefont {M.-W.}\ \bibnamefont {Chang}}, \bibinfo {author} {\bibfnamefont {K.}~\bibnamefont {Lee}},\ and\ \bibinfo {author} {\bibfnamefont {K.}~\bibnamefont {Toutanova}},\ }\href {https://arxiv.org/abs/1810.04805} {\bibinfo {title} {Bert: Pre-training of deep bidirectional transformers for language understanding}} (\bibinfo {year} {2019}),\ \Eprint {https://arxiv.org/abs/1810.04805} {arXiv:1810.04805 [cs.CL]} \BibitemShut {NoStop}%
\bibitem [{\citenamefont {Meignen}\ \emph {et~al.}(2015)\citenamefont {Meignen}, \citenamefont {Gardner},\ and\ \citenamefont {Oberlin}}]{7362631}%
  \BibitemOpen
  \bibfield  {author} {\bibinfo {author} {\bibfnamefont {S.}~\bibnamefont {Meignen}}, \bibinfo {author} {\bibfnamefont {T.}~\bibnamefont {Gardner}},\ and\ \bibinfo {author} {\bibfnamefont {T.}~\bibnamefont {Oberlin}},\ }\bibfield  {title} {\bibinfo {title} {Time-frequency ridge analysis based on the reassignment vector},\ }in\ \href {https://doi.org/10.1109/EUSIPCO.2015.7362631} {\emph {\bibinfo {booktitle} {2015 23rd European Signal Processing Conference (EUSIPCO)}}}\ (\bibinfo {year} {2015})\ pp.\ \bibinfo {pages} {1486--1490}\BibitemShut {NoStop}%
\bibitem [{\citenamefont {Robson}\ \emph {et~al.}(2019{\natexlab{b}})\citenamefont {Robson}, \citenamefont {Cornish},\ and\ \citenamefont {Liu}}]{Robson:2019psd}%
  \BibitemOpen
  \bibfield  {author} {\bibinfo {author} {\bibfnamefont {T.}~\bibnamefont {Robson}}, \bibinfo {author} {\bibfnamefont {N.~J.}\ \bibnamefont {Cornish}},\ and\ \bibinfo {author} {\bibfnamefont {C.~J.}\ \bibnamefont {Liu}},\ }\bibfield  {title} {\bibinfo {title} {{The Construction and Use of LISA Sensitivity Curves}},\ }\href {https://doi.org/10.1088/1361-6382/ab1101} {\bibfield  {journal} {\bibinfo  {journal} {Class.\ Quant.\ Grav.}\ }\textbf {\bibinfo {volume} {36}},\ \bibinfo {pages} {105011} (\bibinfo {year} {2019}{\natexlab{b}})}\BibitemShut {NoStop}%
\bibitem [{\citenamefont {Cornish}\ and\ \citenamefont {Robson}(2017)}]{Cornish_2017}%
  \BibitemOpen
  \bibfield  {author} {\bibinfo {author} {\bibfnamefont {N.}~\bibnamefont {Cornish}}\ and\ \bibinfo {author} {\bibfnamefont {T.}~\bibnamefont {Robson}},\ }\bibfield  {title} {\bibinfo {title} {Galactic binary science with the new lisa design},\ }\href {https://doi.org/10.1088/1742-6596/840/1/012024} {\bibfield  {journal} {\bibinfo  {journal} {Journal of Physics: Conference Series}\ }\textbf {\bibinfo {volume} {840}},\ \bibinfo {pages} {012024} (\bibinfo {year} {2017})}\BibitemShut {NoStop}%
\bibitem [{\citenamefont {Baibhav}\ and\ \citenamefont {Berti}(2019)}]{Baibhav_2019}%
  \BibitemOpen
  \bibfield  {author} {\bibinfo {author} {\bibfnamefont {V.}~\bibnamefont {Baibhav}}\ and\ \bibinfo {author} {\bibfnamefont {E.}~\bibnamefont {Berti}},\ }\bibfield  {title} {\bibinfo {title} {Multimode black hole spectroscopy},\ }\bibfield  {journal} {\bibinfo  {journal} {Physical Review D}\ }\textbf {\bibinfo {volume} {99}},\ \href {https://doi.org/10.1103/physrevd.99.024005} {10.1103/physrevd.99.024005} (\bibinfo {year} {2019})\BibitemShut {NoStop}%
\bibitem [{\citenamefont {Berti}\ \emph {et~al.}(2007)\citenamefont {Berti}, \citenamefont {Cardoso}, \citenamefont {Cardoso},\ and\ \citenamefont {Cavaglià}}]{Berti_2007}%
  \BibitemOpen
  \bibfield  {author} {\bibinfo {author} {\bibfnamefont {E.}~\bibnamefont {Berti}}, \bibinfo {author} {\bibfnamefont {J.}~\bibnamefont {Cardoso}}, \bibinfo {author} {\bibfnamefont {V.}~\bibnamefont {Cardoso}},\ and\ \bibinfo {author} {\bibfnamefont {M.}~\bibnamefont {Cavaglià}},\ }\bibfield  {title} {\bibinfo {title} {Matched filtering and parameter estimation of ringdown waveforms},\ }\bibfield  {journal} {\bibinfo  {journal} {Physical Review D}\ }\textbf {\bibinfo {volume} {76}},\ \href {https://doi.org/10.1103/physrevd.76.104044} {10.1103/physrevd.76.104044} (\bibinfo {year} {2007})\BibitemShut {NoStop}%
\bibitem [{\citenamefont {Berti}\ \emph {et~al.}(2016)\citenamefont {Berti}, \citenamefont {Sesana}, \citenamefont {Barausse}, \citenamefont {Cardoso},\ and\ \citenamefont {Belczynski}}]{Berti_2016}%
  \BibitemOpen
  \bibfield  {author} {\bibinfo {author} {\bibfnamefont {E.}~\bibnamefont {Berti}}, \bibinfo {author} {\bibfnamefont {A.}~\bibnamefont {Sesana}}, \bibinfo {author} {\bibfnamefont {E.}~\bibnamefont {Barausse}}, \bibinfo {author} {\bibfnamefont {V.}~\bibnamefont {Cardoso}},\ and\ \bibinfo {author} {\bibfnamefont {K.}~\bibnamefont {Belczynski}},\ }\bibfield  {title} {\bibinfo {title} {Spectroscopy of kerr black holes with earth- and space-based interferometers},\ }\bibfield  {journal} {\bibinfo  {journal} {Physical Review Letters}\ }\textbf {\bibinfo {volume} {117}},\ \href {https://doi.org/10.1103/physrevlett.117.101102} {10.1103/physrevlett.117.101102} (\bibinfo {year} {2016})\BibitemShut {NoStop}%
\bibitem [{\citenamefont {Barausse}\ and\ \citenamefont {Rezzolla}(2009)}]{Barausse_2009}%
  \BibitemOpen
  \bibfield  {author} {\bibinfo {author} {\bibfnamefont {E.}~\bibnamefont {Barausse}}\ and\ \bibinfo {author} {\bibfnamefont {L.}~\bibnamefont {Rezzolla}},\ }\bibfield  {title} {\bibinfo {title} {Predicting the direction of the final spin from the coalescence of two black holes},\ }\href {https://doi.org/10.1088/0004-637X/704/1/L40} {\bibfield  {journal} {\bibinfo  {journal} {The Astrophysical Journal}\ }\textbf {\bibinfo {volume} {704}},\ \bibinfo {pages} {L40} (\bibinfo {year} {2009})}\BibitemShut {NoStop}%
\bibitem [{\citenamefont {Berti}\ \emph {et~al.}(2006{\natexlab{b}})\citenamefont {Berti}, \citenamefont {Cardoso},\ and\ \citenamefont {Will}}]{Berti_2006}%
  \BibitemOpen
  \bibfield  {author} {\bibinfo {author} {\bibfnamefont {E.}~\bibnamefont {Berti}}, \bibinfo {author} {\bibfnamefont {V.}~\bibnamefont {Cardoso}},\ and\ \bibinfo {author} {\bibfnamefont {C.~M.}\ \bibnamefont {Will}},\ }\bibfield  {title} {\bibinfo {title} {Gravitational-wave spectroscopy of massive black holes with the space interferometer lisa},\ }\bibfield  {journal} {\bibinfo  {journal} {Physical Review D}\ }\textbf {\bibinfo {volume} {73}},\ \href {https://doi.org/10.1103/physrevd.73.064030} {10.1103/physrevd.73.064030} (\bibinfo {year} {2006}{\natexlab{b}})\BibitemShut {NoStop}%
\bibitem [{\citenamefont {Seoane}\ \emph {et~al.}(2021)\citenamefont {Seoane}, \citenamefont {Sedda}, \citenamefont {Babak}, \citenamefont {Berry}, \citenamefont {Berti}, \citenamefont {Bertone}, \citenamefont {Blas}, \citenamefont {Bogdanović}, \citenamefont {Bonetti}, \citenamefont {Breivik} \emph {et~al.}}]{AmaroSeoane:2021}%
  \BibitemOpen
  \bibfield  {author} {\bibinfo {author} {\bibfnamefont {P.~A.}\ \bibnamefont {Seoane}}, \bibinfo {author} {\bibfnamefont {M.~A.}\ \bibnamefont {Sedda}}, \bibinfo {author} {\bibfnamefont {S.}~\bibnamefont {Babak}}, \bibinfo {author} {\bibfnamefont {C.~P.~L.}\ \bibnamefont {Berry}}, \bibinfo {author} {\bibfnamefont {E.}~\bibnamefont {Berti}}, \bibinfo {author} {\bibfnamefont {G.}~\bibnamefont {Bertone}}, \bibinfo {author} {\bibfnamefont {D.}~\bibnamefont {Blas}}, \bibinfo {author} {\bibfnamefont {T.}~\bibnamefont {Bogdanović}}, \bibinfo {author} {\bibfnamefont {M.}~\bibnamefont {Bonetti}}, \bibinfo {author} {\bibfnamefont {K.}~\bibnamefont {Breivik}}, \emph {et~al.},\ }\bibfield  {title} {\bibinfo {title} {Astrophysics with the laser interferometer space antenna},\ }\href {https://doi.org/10.1007/s10714-021-02889-x} {\bibfield  {journal} {\bibinfo  {journal} {General Relativity and Gravitation}\ }\textbf {\bibinfo {volume} {54}},\ \bibinfo {pages} {3} (\bibinfo {year} {2021})}\BibitemShut {NoStop}%
\bibitem [{\citenamefont {Harris}(1978{\natexlab{b}})}]{Harris:1978}%
  \BibitemOpen
  \bibfield  {author} {\bibinfo {author} {\bibfnamefont {F.~J.}\ \bibnamefont {Harris}},\ }\bibfield  {title} {\bibinfo {title} {On the use of windows for harmonic analysis with the discrete fourier transform},\ }\href {https://doi.org/10.1109/PROC.1978.10837} {\bibfield  {journal} {\bibinfo  {journal} {Proceedings of the IEEE}\ }\textbf {\bibinfo {volume} {66}},\ \bibinfo {pages} {51} (\bibinfo {year} {1978}{\natexlab{b}})}\BibitemShut {NoStop}%
\bibitem [{\citenamefont {McNabb}\ \emph {et~al.}(2019)\citenamefont {McNabb}, \citenamefont {Capano},\ and\ \citenamefont {Hanna}}]{McNabb:2019spectrogram}%
  \BibitemOpen
  \bibfield  {author} {\bibinfo {author} {\bibfnamefont {J.}~\bibnamefont {McNabb}}, \bibinfo {author} {\bibfnamefont {C.~D.}\ \bibnamefont {Capano}},\ and\ \bibinfo {author} {\bibfnamefont {C.~R.}\ \bibnamefont {Hanna}},\ }\bibfield  {title} {\bibinfo {title} {{Spectrogram-Based Searches for Ringdown Signals}},\ }\href {https://doi.org/10.1103/PhysRevD.100.123010} {\bibfield  {journal} {\bibinfo  {journal} {Phys.\ Rev.\ D}\ }\textbf {\bibinfo {volume} {100}},\ \bibinfo {pages} {123010} (\bibinfo {year} {2019})}\BibitemShut {NoStop}%
\bibitem [{\citenamefont {Mallat}(1999)}]{MALLAT199942}%
  \BibitemOpen
  \bibfield  {author} {\bibinfo {author} {\bibfnamefont {S.}~\bibnamefont {Mallat}},\ }\bibfield  {title} {\bibinfo {title} {Iii - discrete revolution},\ }in\ \href {https://doi.org/https://doi.org/10.1016/B978-012466606-1/50005-2} {\emph {\bibinfo {booktitle} {A Wavelet Tour of Signal Processing (Second Edition)}}},\ \bibinfo {editor} {edited by\ \bibinfo {editor} {\bibfnamefont {S.}~\bibnamefont {Mallat}}}\ (\bibinfo  {publisher} {Academic Press},\ \bibinfo {address} {San Diego},\ \bibinfo {year} {1999})\ \bibinfo {edition} {second edition}\ ed.,\ pp.\ \bibinfo {pages} {42--66}\BibitemShut {NoStop}%
\bibitem [{\citenamefont {Younesi}\ \emph {et~al.}(2024)\citenamefont {Younesi}, \citenamefont {Ansari}, \citenamefont {Fazli}, \citenamefont {Ejlali}, \citenamefont {Shafique},\ and\ \citenamefont {Henkel}}]{younesi2024comprehensivesurveyconvolutionsdeep}%
  \BibitemOpen
  \bibfield  {author} {\bibinfo {author} {\bibfnamefont {A.}~\bibnamefont {Younesi}}, \bibinfo {author} {\bibfnamefont {M.}~\bibnamefont {Ansari}}, \bibinfo {author} {\bibfnamefont {M.}~\bibnamefont {Fazli}}, \bibinfo {author} {\bibfnamefont {A.}~\bibnamefont {Ejlali}}, \bibinfo {author} {\bibfnamefont {M.}~\bibnamefont {Shafique}},\ and\ \bibinfo {author} {\bibfnamefont {J.}~\bibnamefont {Henkel}},\ }\href {https://arxiv.org/abs/2402.15490} {\bibinfo {title} {A comprehensive survey of convolutions in deep learning: Applications, challenges, and future trends}} (\bibinfo {year} {2024}),\ \Eprint {https://arxiv.org/abs/2402.15490} {arXiv:2402.15490 [cs.LG]} \BibitemShut {NoStop}%
\bibitem [{\citenamefont {Vaswani}\ \emph {et~al.}(2023)\citenamefont {Vaswani}, \citenamefont {Shazeer}, \citenamefont {Parmar}, \citenamefont {Uszkoreit}, \citenamefont {Jones}, \citenamefont {Gomez}, \citenamefont {Kaiser},\ and\ \citenamefont {Polosukhin}}]{vaswani2023attentionneed}%
  \BibitemOpen
  \bibfield  {author} {\bibinfo {author} {\bibfnamefont {A.}~\bibnamefont {Vaswani}}, \bibinfo {author} {\bibfnamefont {N.}~\bibnamefont {Shazeer}}, \bibinfo {author} {\bibfnamefont {N.}~\bibnamefont {Parmar}}, \bibinfo {author} {\bibfnamefont {J.}~\bibnamefont {Uszkoreit}}, \bibinfo {author} {\bibfnamefont {L.}~\bibnamefont {Jones}}, \bibinfo {author} {\bibfnamefont {A.~N.}\ \bibnamefont {Gomez}}, \bibinfo {author} {\bibfnamefont {L.}~\bibnamefont {Kaiser}},\ and\ \bibinfo {author} {\bibfnamefont {I.}~\bibnamefont {Polosukhin}},\ }\href {https://arxiv.org/abs/1706.03762} {\bibinfo {title} {Attention is all you need}} (\bibinfo {year} {2023}),\ \Eprint {https://arxiv.org/abs/1706.03762} {arXiv:1706.03762 [cs.CL]} \BibitemShut {NoStop}%
\bibitem [{\citenamefont {Nie}\ \emph {et~al.}(2023)\citenamefont {Nie}, \citenamefont {Nguyen}, \citenamefont {Sinthong},\ and\ \citenamefont {Kalagnanam}}]{nie2023timeseriesworth64}%
  \BibitemOpen
  \bibfield  {author} {\bibinfo {author} {\bibfnamefont {Y.}~\bibnamefont {Nie}}, \bibinfo {author} {\bibfnamefont {N.~H.}\ \bibnamefont {Nguyen}}, \bibinfo {author} {\bibfnamefont {P.}~\bibnamefont {Sinthong}},\ and\ \bibinfo {author} {\bibfnamefont {J.}~\bibnamefont {Kalagnanam}},\ }\href {https://arxiv.org/abs/2211.14730} {\bibinfo {title} {A time series is worth 64 words: Long-term forecasting with transformers}} (\bibinfo {year} {2023}),\ \Eprint {https://arxiv.org/abs/2211.14730} {arXiv:2211.14730 [cs.LG]} \BibitemShut {NoStop}%
\bibitem [{\citenamefont {Xiong}\ \emph {et~al.}(2020)\citenamefont {Xiong}, \citenamefont {Yang}, \citenamefont {He}, \citenamefont {Zheng}, \citenamefont {Zheng}, \citenamefont {Xing}, \citenamefont {Zhang}, \citenamefont {Lan}, \citenamefont {Wang},\ and\ \citenamefont {Liu}}]{xiong2020layernormalizationtransformerarchitecture}%
  \BibitemOpen
  \bibfield  {author} {\bibinfo {author} {\bibfnamefont {R.}~\bibnamefont {Xiong}}, \bibinfo {author} {\bibfnamefont {Y.}~\bibnamefont {Yang}}, \bibinfo {author} {\bibfnamefont {D.}~\bibnamefont {He}}, \bibinfo {author} {\bibfnamefont {K.}~\bibnamefont {Zheng}}, \bibinfo {author} {\bibfnamefont {S.}~\bibnamefont {Zheng}}, \bibinfo {author} {\bibfnamefont {C.}~\bibnamefont {Xing}}, \bibinfo {author} {\bibfnamefont {H.}~\bibnamefont {Zhang}}, \bibinfo {author} {\bibfnamefont {Y.}~\bibnamefont {Lan}}, \bibinfo {author} {\bibfnamefont {L.}~\bibnamefont {Wang}},\ and\ \bibinfo {author} {\bibfnamefont {T.-Y.}\ \bibnamefont {Liu}},\ }\href {https://arxiv.org/abs/2002.04745} {\bibinfo {title} {On layer normalization in the transformer architecture}} (\bibinfo {year} {2020}),\ \Eprint {https://arxiv.org/abs/2002.04745} {arXiv:2002.04745 [cs.LG]} \BibitemShut {NoStop}%
\bibitem [{\citenamefont {Mao}\ \emph {et~al.}(2024)\citenamefont {Mao}, \citenamefont {Lee},\ and\ \citenamefont {Edwards}}]{mao2024bigru}%
  \BibitemOpen
  \bibfield  {author} {\bibinfo {author} {\bibfnamefont {R.}~\bibnamefont {Mao}}, \bibinfo {author} {\bibfnamefont {J.~E.}\ \bibnamefont {Lee}},\ and\ \bibinfo {author} {\bibfnamefont {M.~C.}\ \bibnamefont {Edwards}},\ }\bibfield  {title} {\bibinfo {title} {A novel stacked hybrid autoencoder for imputing lisa data gaps},\ }\href@noop {} {\bibfield  {journal} {\bibinfo  {journal} {arXiv preprint arXiv:2410.05571}\ } (\bibinfo {year} {2024})}\BibitemShut {NoStop}%
\bibitem [{\citenamefont {Zhang}\ \emph {et~al.}(2018{\natexlab{a}})\citenamefont {Zhang}, \citenamefont {Isola}, \citenamefont {Efros}, \citenamefont {Shechtman},\ and\ \citenamefont {Wang}}]{zhang2018unreasonableeffectivenessdeepfeatures}%
  \BibitemOpen
  \bibfield  {author} {\bibinfo {author} {\bibfnamefont {R.}~\bibnamefont {Zhang}}, \bibinfo {author} {\bibfnamefont {P.}~\bibnamefont {Isola}}, \bibinfo {author} {\bibfnamefont {A.~A.}\ \bibnamefont {Efros}}, \bibinfo {author} {\bibfnamefont {E.}~\bibnamefont {Shechtman}},\ and\ \bibinfo {author} {\bibfnamefont {O.}~\bibnamefont {Wang}},\ }\href {https://arxiv.org/abs/1801.03924} {\bibinfo {title} {The unreasonable effectiveness of deep features as a perceptual metric}} (\bibinfo {year} {2018}{\natexlab{a}}),\ \Eprint {https://arxiv.org/abs/1801.03924} {arXiv:1801.03924 [cs.CV]} \BibitemShut {NoStop}%
\bibitem [{\citenamefont {Zhang}\ \emph {et~al.}(2021)\citenamefont {Zhang}, \citenamefont {Gong}, \citenamefont {Zhang} \emph {et~al.}}]{zhang2021parameter}%
  \BibitemOpen
  \bibfield  {author} {\bibinfo {author} {\bibfnamefont {C.}~\bibnamefont {Zhang}}, \bibinfo {author} {\bibfnamefont {Y.}~\bibnamefont {Gong}}, \bibinfo {author} {\bibfnamefont {C.}~\bibnamefont {Zhang}}, \emph {et~al.},\ }\bibfield  {title} {\bibinfo {title} {Parameter estimation for space-based gravitational wave detectors with ringdown signals},\ }\href@noop {} {\bibfield  {journal} {\bibinfo  {journal} {Physical Review D}\ }\textbf {\bibinfo {volume} {104}},\ \bibinfo {pages} {083038} (\bibinfo {year} {2021})}\BibitemShut {NoStop}%
\bibitem [{\citenamefont {Babak}\ \emph {et~al.}(2008)\citenamefont {Babak}, \citenamefont {Baker}, \citenamefont {Benacquista}, \citenamefont {Cornish}, \citenamefont {Crowder}, \citenamefont {Larson}, \citenamefont {Plagnol}, \citenamefont {Porter}, \citenamefont {Vallisneri}, \citenamefont {Vecchio}, \citenamefont {Arnaud}, \citenamefont {Barack}, \citenamefont {Błaut}, \citenamefont {Cutler}, \citenamefont {Fairhurst}, \citenamefont {Gair}, \citenamefont {Gong}, \citenamefont {Harry}, \citenamefont {Khurana}, \citenamefont {Królak}, \citenamefont {Mandel}, \citenamefont {Prix}, \citenamefont {Sathyaprakash}, \citenamefont {Savov}, \citenamefont {Shang}, \citenamefont {Trias}, \citenamefont {Veitch}, \citenamefont {Wang}, \citenamefont {Wen},\ and\ \citenamefont {T~Whelan}}]{Babak_2008}%
  \BibitemOpen
  \bibfield  {author} {\bibinfo {author} {\bibfnamefont {S.}~\bibnamefont {Babak}}, \bibinfo {author} {\bibfnamefont {J.~G.}\ \bibnamefont {Baker}}, \bibinfo {author} {\bibfnamefont {M.~J.}\ \bibnamefont {Benacquista}}, \bibinfo {author} {\bibfnamefont {N.~J.}\ \bibnamefont {Cornish}}, \bibinfo {author} {\bibfnamefont {J.}~\bibnamefont {Crowder}}, \bibinfo {author} {\bibfnamefont {S.~L.}\ \bibnamefont {Larson}}, \bibinfo {author} {\bibfnamefont {E.}~\bibnamefont {Plagnol}}, \bibinfo {author} {\bibfnamefont {E.~K.}\ \bibnamefont {Porter}}, \bibinfo {author} {\bibfnamefont {M.}~\bibnamefont {Vallisneri}}, \bibinfo {author} {\bibfnamefont {A.}~\bibnamefont {Vecchio}}, \bibinfo {author} {\bibfnamefont {K.}~\bibnamefont {Arnaud}}, \bibinfo {author} {\bibfnamefont {L.}~\bibnamefont {Barack}}, \bibinfo {author} {\bibfnamefont {A.}~\bibnamefont {Błaut}}, \bibinfo {author} {\bibfnamefont {C.}~\bibnamefont {Cutler}}, \bibinfo {author} {\bibfnamefont {S.}~\bibnamefont {Fairhurst}}, \bibinfo {author} {\bibfnamefont
  {J.}~\bibnamefont {Gair}}, \bibinfo {author} {\bibfnamefont {X.}~\bibnamefont {Gong}}, \bibinfo {author} {\bibfnamefont {I.}~\bibnamefont {Harry}}, \bibinfo {author} {\bibfnamefont {D.}~\bibnamefont {Khurana}}, \bibinfo {author} {\bibfnamefont {A.}~\bibnamefont {Królak}}, \bibinfo {author} {\bibfnamefont {I.}~\bibnamefont {Mandel}}, \bibinfo {author} {\bibfnamefont {R.}~\bibnamefont {Prix}}, \bibinfo {author} {\bibfnamefont {B.~S.}\ \bibnamefont {Sathyaprakash}}, \bibinfo {author} {\bibfnamefont {P.}~\bibnamefont {Savov}}, \bibinfo {author} {\bibfnamefont {Y.}~\bibnamefont {Shang}}, \bibinfo {author} {\bibfnamefont {M.}~\bibnamefont {Trias}}, \bibinfo {author} {\bibfnamefont {J.}~\bibnamefont {Veitch}}, \bibinfo {author} {\bibfnamefont {Y.}~\bibnamefont {Wang}}, \bibinfo {author} {\bibfnamefont {L.}~\bibnamefont {Wen}},\ and\ \bibinfo {author} {\bibfnamefont {J.}~\bibnamefont {T~Whelan}},\ }\bibfield  {title} {\bibinfo {title} {The mock lisa data challenges: from challenge 1b to challenge 3},\ }\href
  {https://doi.org/10.1088/0264-9381/25/18/184026} {\bibfield  {journal} {\bibinfo  {journal} {Classical and Quantum Gravity}\ }\textbf {\bibinfo {volume} {25}},\ \bibinfo {pages} {184026} (\bibinfo {year} {2008})}\BibitemShut {NoStop}%
\bibitem [{\citenamefont {Babak}\ \emph {et~al.}(2010)\citenamefont {Babak}, \citenamefont {Baker}, \citenamefont {Benacquista}, \citenamefont {Cornish}, \citenamefont {Larson}, \citenamefont {Mandel}, \citenamefont {McWilliams}, \citenamefont {Petiteau}, \citenamefont {Porter}, \citenamefont {Robinson}, \citenamefont {Vallisneri}, \citenamefont {Vecchio}, \citenamefont {Adams}, \citenamefont {Arnaud}, \citenamefont {Błaut}, \citenamefont {Bridges}, \citenamefont {Cohen}, \citenamefont {Cutler}, \citenamefont {Feroz}, \citenamefont {Gair}, \citenamefont {Graff}, \citenamefont {Hobson}, \citenamefont {Key}, \citenamefont {Królak}, \citenamefont {Lasenby}, \citenamefont {Prix}, \citenamefont {Shang}, \citenamefont {Trias}, \citenamefont {Veitch},\ and\ \citenamefont {Whelan}}]{Babak_2010}%
  \BibitemOpen
  \bibfield  {author} {\bibinfo {author} {\bibfnamefont {S.}~\bibnamefont {Babak}}, \bibinfo {author} {\bibfnamefont {J.~G.}\ \bibnamefont {Baker}}, \bibinfo {author} {\bibfnamefont {M.~J.}\ \bibnamefont {Benacquista}}, \bibinfo {author} {\bibfnamefont {N.~J.}\ \bibnamefont {Cornish}}, \bibinfo {author} {\bibfnamefont {S.~L.}\ \bibnamefont {Larson}}, \bibinfo {author} {\bibfnamefont {I.}~\bibnamefont {Mandel}}, \bibinfo {author} {\bibfnamefont {S.~T.}\ \bibnamefont {McWilliams}}, \bibinfo {author} {\bibfnamefont {A.}~\bibnamefont {Petiteau}}, \bibinfo {author} {\bibfnamefont {E.~K.}\ \bibnamefont {Porter}}, \bibinfo {author} {\bibfnamefont {E.~L.}\ \bibnamefont {Robinson}}, \bibinfo {author} {\bibfnamefont {M.}~\bibnamefont {Vallisneri}}, \bibinfo {author} {\bibfnamefont {A.}~\bibnamefont {Vecchio}}, \bibinfo {author} {\bibfnamefont {M.}~\bibnamefont {Adams}}, \bibinfo {author} {\bibfnamefont {K.~A.}\ \bibnamefont {Arnaud}}, \bibinfo {author} {\bibfnamefont {A.}~\bibnamefont {Błaut}}, \bibinfo {author}
  {\bibfnamefont {M.}~\bibnamefont {Bridges}}, \bibinfo {author} {\bibfnamefont {M.}~\bibnamefont {Cohen}}, \bibinfo {author} {\bibfnamefont {C.}~\bibnamefont {Cutler}}, \bibinfo {author} {\bibfnamefont {F.}~\bibnamefont {Feroz}}, \bibinfo {author} {\bibfnamefont {J.~R.}\ \bibnamefont {Gair}}, \bibinfo {author} {\bibfnamefont {P.}~\bibnamefont {Graff}}, \bibinfo {author} {\bibfnamefont {M.}~\bibnamefont {Hobson}}, \bibinfo {author} {\bibfnamefont {J.~S.}\ \bibnamefont {Key}}, \bibinfo {author} {\bibfnamefont {A.}~\bibnamefont {Królak}}, \bibinfo {author} {\bibfnamefont {A.}~\bibnamefont {Lasenby}}, \bibinfo {author} {\bibfnamefont {R.}~\bibnamefont {Prix}}, \bibinfo {author} {\bibfnamefont {Y.}~\bibnamefont {Shang}}, \bibinfo {author} {\bibfnamefont {M.}~\bibnamefont {Trias}}, \bibinfo {author} {\bibfnamefont {J.}~\bibnamefont {Veitch}},\ and\ \bibinfo {author} {\bibfnamefont {J.~T.}\ \bibnamefont {Whelan}},\ }\bibfield  {title} {\bibinfo {title} {The mock lisa data challenges: from challenge 3 to challenge
  4},\ }\href {https://doi.org/10.1088/0264-9381/27/8/084009} {\bibfield  {journal} {\bibinfo  {journal} {Classical and Quantum Gravity}\ }\textbf {\bibinfo {volume} {27}},\ \bibinfo {pages} {084009} (\bibinfo {year} {2010})}\BibitemShut {NoStop}%
\bibitem [{\citenamefont {Berti}(2022)}]{Berti:2022kcw}%
  \BibitemOpen
  \bibfield  {author} {\bibinfo {author} {\bibfnamefont {E.}~\bibnamefont {Berti}},\ }\bibfield  {title} {\bibinfo {title} {{Black Hole Spectroscopy and Quasinormal Mode Catalogs}},\ }\href {https://doi.org/10.1088/1361-6382/acb083} {\bibfield  {journal} {\bibinfo  {journal} {Class.\ Quant.\ Grav.}\ }\textbf {\bibinfo {volume} {39}},\ \bibinfo {pages} {244001} (\bibinfo {year} {2022})}\BibitemShut {NoStop}%
\bibitem [{\citenamefont {Robson}\ \emph {et~al.}(2019{\natexlab{c}})\citenamefont {Robson}, \citenamefont {Cornish},\ and\ \citenamefont {Liu}}]{Robson_2019}%
  \BibitemOpen
  \bibfield  {author} {\bibinfo {author} {\bibfnamefont {T.}~\bibnamefont {Robson}}, \bibinfo {author} {\bibfnamefont {N.~J.}\ \bibnamefont {Cornish}},\ and\ \bibinfo {author} {\bibfnamefont {C.}~\bibnamefont {Liu}},\ }\bibfield  {title} {\bibinfo {title} {The construction and use of lisa sensitivity curves},\ }\href {https://doi.org/10.1088/1361-6382/ab1101} {\bibfield  {journal} {\bibinfo  {journal} {Classical and Quantum Gravity}\ }\textbf {\bibinfo {volume} {36}},\ \bibinfo {pages} {105011} (\bibinfo {year} {2019}{\natexlab{c}})}\BibitemShut {NoStop}%
\bibitem [{\citenamefont {Khan}\ \emph {et~al.}(2016)\citenamefont {Khan}, \citenamefont {Husa}, \citenamefont {Hannam}, \citenamefont {Ohme}, \citenamefont {Pürrer}, \citenamefont {Forteza},\ and\ \citenamefont {Boh{\'e}}}]{Khan_2016}%
  \BibitemOpen
  \bibfield  {author} {\bibinfo {author} {\bibfnamefont {S.}~\bibnamefont {Khan}}, \bibinfo {author} {\bibfnamefont {S.}~\bibnamefont {Husa}}, \bibinfo {author} {\bibfnamefont {M.}~\bibnamefont {Hannam}}, \bibinfo {author} {\bibfnamefont {F.}~\bibnamefont {Ohme}}, \bibinfo {author} {\bibfnamefont {M.}~\bibnamefont {Pürrer}}, \bibinfo {author} {\bibfnamefont {X.~J.}\ \bibnamefont {Forteza}},\ and\ \bibinfo {author} {\bibfnamefont {A.}~\bibnamefont {Boh{\'e}}},\ }\bibfield  {title} {\bibinfo {title} {Frequency-domain gravitational waves from nonprecessing black-hole binaries. ii. a phenomenological model for the advanced detector era},\ }\bibfield  {journal} {\bibinfo  {journal} {Physical Review D}\ }\textbf {\bibinfo {volume} {93}},\ \href {https://doi.org/10.1103/physrevd.93.044007} {10.1103/physrevd.93.044007} (\bibinfo {year} {2016})\BibitemShut {NoStop}%
\bibitem [{\citenamefont {Husa}\ \emph {et~al.}(2016)\citenamefont {Husa}, \citenamefont {Khan}, \citenamefont {Hannam}, \citenamefont {Pürrer}, \citenamefont {Ohme}, \citenamefont {Forteza},\ and\ \citenamefont {Boh{\'e}}}]{Husa_2016}%
  \BibitemOpen
  \bibfield  {author} {\bibinfo {author} {\bibfnamefont {S.}~\bibnamefont {Husa}}, \bibinfo {author} {\bibfnamefont {S.}~\bibnamefont {Khan}}, \bibinfo {author} {\bibfnamefont {M.}~\bibnamefont {Hannam}}, \bibinfo {author} {\bibfnamefont {M.}~\bibnamefont {Pürrer}}, \bibinfo {author} {\bibfnamefont {F.}~\bibnamefont {Ohme}}, \bibinfo {author} {\bibfnamefont {X.~J.}\ \bibnamefont {Forteza}},\ and\ \bibinfo {author} {\bibfnamefont {A.}~\bibnamefont {Boh{\'e}}},\ }\bibfield  {title} {\bibinfo {title} {Frequency-domain gravitational waves from nonprecessing black-hole binaries. i. new numerical waveforms and anatomy of the signal},\ }\bibfield  {journal} {\bibinfo  {journal} {Physical Review D}\ }\textbf {\bibinfo {volume} {93}},\ \href {https://doi.org/10.1103/physrevd.93.044006} {10.1103/physrevd.93.044006} (\bibinfo {year} {2016})\BibitemShut {NoStop}%
\bibitem [{\citenamefont {Cutler}\ and\ \citenamefont {Flanagan}(1994)}]{Cutler_1994}%
  \BibitemOpen
  \bibfield  {author} {\bibinfo {author} {\bibfnamefont {C.}~\bibnamefont {Cutler}}\ and\ \bibinfo {author} {\bibfnamefont {{\'E}.~E.}\ \bibnamefont {Flanagan}},\ }\bibfield  {title} {\bibinfo {title} {Gravitational waves from merging compact binaries: How accurately can one extract the binary’s parameters from the inspiral waveform?},\ }\href {https://doi.org/10.1103/physrevd.49.2658} {\bibfield  {journal} {\bibinfo  {journal} {Physical Review D}\ }\textbf {\bibinfo {volume} {49}},\ \bibinfo {pages} {2658–2697} (\bibinfo {year} {1994})}\BibitemShut {NoStop}%
\bibitem [{\citenamefont {Owen}(1996)}]{Owen:1996PRD}%
  \BibitemOpen
  \bibfield  {author} {\bibinfo {author} {\bibfnamefont {B.~J.}\ \bibnamefont {Owen}},\ }\bibfield  {title} {\bibinfo {title} {Search templates for gravitational waves from inspiraling binaries: Choice of template spacing},\ }\href {https://doi.org/10.1103/PhysRevD.53.6749} {\bibfield  {journal} {\bibinfo  {journal} {Physical Review D}\ }\textbf {\bibinfo {volume} {53}},\ \bibinfo {pages} {6749} (\bibinfo {year} {1996})},\ \Eprint {https://arxiv.org/abs/gr-qc/9511032} {arXiv:gr-qc/9511032} \BibitemShut {NoStop}%
\bibitem [{\citenamefont {Thrane}\ and\ \citenamefont {Talbot}(2019)}]{Thrane:2019PASA}%
  \BibitemOpen
  \bibfield  {author} {\bibinfo {author} {\bibfnamefont {E.}~\bibnamefont {Thrane}}\ and\ \bibinfo {author} {\bibfnamefont {C.}~\bibnamefont {Talbot}},\ }\bibfield  {title} {\bibinfo {title} {An introduction to bayesian inference in gravitational-wave astronomy: parameter estimation, model selection, and hierarchical models},\ }\href {https://doi.org/10.1017/pasa.2019.2} {\bibfield  {journal} {\bibinfo  {journal} {Publications of the Astronomical Society of Australia}\ }\textbf {\bibinfo {volume} {36}},\ \bibinfo {pages} {e010} (\bibinfo {year} {2019})},\ \Eprint {https://arxiv.org/abs/1809.02293} {arXiv:1809.02293} \BibitemShut {NoStop}%
\bibitem [{\citenamefont {Veitch}\ \emph {et~al.}(2015)\citenamefont {Veitch} \emph {et~al.}}]{Veitch:2015PRD}%
  \BibitemOpen
  \bibfield  {author} {\bibinfo {author} {\bibfnamefont {J.}~\bibnamefont {Veitch}} \emph {et~al.},\ }\bibfield  {title} {\bibinfo {title} {Parameter estimation for compact binaries with ground-based gravitational-wave observations using the {LALInference} software library},\ }\href {https://doi.org/10.1103/PhysRevD.91.042003} {\bibfield  {journal} {\bibinfo  {journal} {Physical Review D}\ }\textbf {\bibinfo {volume} {91}},\ \bibinfo {pages} {042003} (\bibinfo {year} {2015})},\ \Eprint {https://arxiv.org/abs/1409.7215} {arXiv:1409.7215} \BibitemShut {NoStop}%
\bibitem [{\citenamefont {Baghi}\ \emph {et~al.}(2019{\natexlab{c}})\citenamefont {Baghi}, \citenamefont {Thorpe}, \citenamefont {Slutsky}, \citenamefont {Baker}, \citenamefont {Canton}, \citenamefont {Korsakova},\ and\ \citenamefont {Karnesis}}]{Baghi_2019}%
  \BibitemOpen
  \bibfield  {author} {\bibinfo {author} {\bibfnamefont {Q.}~\bibnamefont {Baghi}}, \bibinfo {author} {\bibfnamefont {J.~I.}\ \bibnamefont {Thorpe}}, \bibinfo {author} {\bibfnamefont {J.}~\bibnamefont {Slutsky}}, \bibinfo {author} {\bibfnamefont {J.}~\bibnamefont {Baker}}, \bibinfo {author} {\bibfnamefont {T.~D.}\ \bibnamefont {Canton}}, \bibinfo {author} {\bibfnamefont {N.}~\bibnamefont {Korsakova}},\ and\ \bibinfo {author} {\bibfnamefont {N.}~\bibnamefont {Karnesis}},\ }\bibfield  {title} {\bibinfo {title} {Gravitational-wave parameter estimation with gaps in lisa: A bayesian data augmentation method},\ }\bibfield  {journal} {\bibinfo  {journal} {Physical Review D}\ }\textbf {\bibinfo {volume} {100}},\ \href {https://doi.org/10.1103/physrevd.100.022003} {10.1103/physrevd.100.022003} (\bibinfo {year} {2019}{\natexlab{c}})\BibitemShut {NoStop}%
\bibitem [{\citenamefont {Zhang}\ \emph {et~al.}(2018{\natexlab{b}})\citenamefont {Zhang}, \citenamefont {Isola}, \citenamefont {Efros}, \citenamefont {Shechtman},\ and\ \citenamefont {Wang}}]{zhang2018unreasonable}%
  \BibitemOpen
  \bibfield  {author} {\bibinfo {author} {\bibfnamefont {R.}~\bibnamefont {Zhang}}, \bibinfo {author} {\bibfnamefont {P.}~\bibnamefont {Isola}}, \bibinfo {author} {\bibfnamefont {A.~A.}\ \bibnamefont {Efros}}, \bibinfo {author} {\bibfnamefont {E.}~\bibnamefont {Shechtman}},\ and\ \bibinfo {author} {\bibfnamefont {O.}~\bibnamefont {Wang}},\ }\bibfield  {title} {\bibinfo {title} {The unreasonable effectiveness of deep features as a perceptual metric},\ }in\ \href@noop {} {\emph {\bibinfo {booktitle} {Proceedings of the IEEE conference on computer vision and pattern recognition}}}\ (\bibinfo {year} {2018})\ pp.\ \bibinfo {pages} {586--595}\BibitemShut {NoStop}%
\bibitem [{\citenamefont {Chatterji}\ \emph {et~al.}(2004{\natexlab{b}})\citenamefont {Chatterji}, \citenamefont {Blackburn}, \citenamefont {Martin},\ and\ \citenamefont {Katsavounidis}}]{Chatterji:2004}%
  \BibitemOpen
  \bibfield  {author} {\bibinfo {author} {\bibfnamefont {S.}~\bibnamefont {Chatterji}}, \bibinfo {author} {\bibfnamefont {L.}~\bibnamefont {Blackburn}}, \bibinfo {author} {\bibfnamefont {G.}~\bibnamefont {Martin}},\ and\ \bibinfo {author} {\bibfnamefont {E.}~\bibnamefont {Katsavounidis}},\ }\bibfield  {title} {\bibinfo {title} {Multiresolution techniques for the detection of gravitational-wave bursts},\ }\href {https://doi.org/10.1088/0264-9381/21/20/024} {\bibfield  {journal} {\bibinfo  {journal} {Classical and Quantum Gravity}\ }\textbf {\bibinfo {volume} {21}},\ \bibinfo {pages} {S1809} (\bibinfo {year} {2004}{\natexlab{b}})}\BibitemShut {NoStop}%
\bibitem [{\citenamefont {Berti}\ \emph {et~al.}(2009{\natexlab{b}})\citenamefont {Berti}, \citenamefont {Cardoso},\ and\ \citenamefont {Starinets}}]{Berti:2009}%
  \BibitemOpen
  \bibfield  {author} {\bibinfo {author} {\bibfnamefont {E.}~\bibnamefont {Berti}}, \bibinfo {author} {\bibfnamefont {V.}~\bibnamefont {Cardoso}},\ and\ \bibinfo {author} {\bibfnamefont {A.~O.}\ \bibnamefont {Starinets}},\ }\bibfield  {title} {\bibinfo {title} {Quasinormal modes of black holes and black branes},\ }\href {https://doi.org/10.1088/0264-9381/26/16/163001} {\bibfield  {journal} {\bibinfo  {journal} {Classical and Quantum Gravity}\ }\textbf {\bibinfo {volume} {26}},\ \bibinfo {pages} {163001} (\bibinfo {year} {2009}{\natexlab{b}})}\BibitemShut {NoStop}%
\bibitem [{\citenamefont {Boashash}(1992)}]{135376}%
  \BibitemOpen
  \bibfield  {author} {\bibinfo {author} {\bibfnamefont {B.}~\bibnamefont {Boashash}},\ }\bibfield  {title} {\bibinfo {title} {Estimating and interpreting the instantaneous frequency of a signal. i. fundamentals},\ }\href {https://doi.org/10.1109/5.135376} {\bibfield  {journal} {\bibinfo  {journal} {Proceedings of the IEEE}\ }\textbf {\bibinfo {volume} {80}},\ \bibinfo {pages} {520} (\bibinfo {year} {1992})}\BibitemShut {NoStop}%
\bibitem [{\citenamefont {Bretthorst}(1988)}]{Bretthorst1988}%
  \BibitemOpen
  \bibfield  {author} {\bibinfo {author} {\bibfnamefont {G.~L.}\ \bibnamefont {Bretthorst}},\ }\href {https://doi.org/10.1007/978-3-642-97190-9} {\emph {\bibinfo {title} {Bayesian Spectrum Analysis and Parameter Estimation}}}\ (\bibinfo  {publisher} {Springer},\ \bibinfo {address} {Berlin},\ \bibinfo {year} {1988})\BibitemShut {NoStop}%
\bibitem [{\citenamefont {Xu}\ \emph {et~al.}(2024{\natexlab{b}})\citenamefont {Xu}, \citenamefont {Du}, \citenamefont {Xu}, \citenamefont {Liang},\ and\ \citenamefont {Wang}}]{xu2024dense}%
  \BibitemOpen
  \bibfield  {author} {\bibinfo {author} {\bibfnamefont {Y.}~\bibnamefont {Xu}}, \bibinfo {author} {\bibfnamefont {M.}~\bibnamefont {Du}}, \bibinfo {author} {\bibfnamefont {P.}~\bibnamefont {Xu}}, \bibinfo {author} {\bibfnamefont {B.}~\bibnamefont {Liang}},\ and\ \bibinfo {author} {\bibfnamefont {H.}~\bibnamefont {Wang}},\ }\bibfield  {title} {\bibinfo {title} {Deep learning enhanced detection of coalescence signals with gap and glitch contamination},\ }\href@noop {} {\bibfield  {journal} {\bibinfo  {journal} {Physical Review D}\ }\textbf {\bibinfo {volume} {109}},\ \bibinfo {pages} {043014} (\bibinfo {year} {2024}{\natexlab{b}})}\BibitemShut {NoStop}%
\end{thebibliography}%

\end{document}